\pdfoutput=1
\documentclass{./style/JINST}
\usepackage{graphicx}
\usepackage{caption}
\usepackage{subcaption}
\usepackage{url}
\usepackage{lineno}
\graphicspath{{./figures/}}

\title{Performance of a segmented $^{6}$Li-loaded liquid scintillator
  detector for the PROSPECT experiment}

\author{The PROSPECT Collaboration}
\author{
J. Ashenfelter$^{p}$,
A. B. Balantekin$^{m}$,
H. R. Band$^{p}$,
C. D. Bass$^{f}$,
D. E. Bergeron$^{g}$,
D. Berish$^{j}$,
N. S. Bowden$^{e}$,
J. P. Brodsky$^{e}$,
C. D. Bryan$^{h}$,
A. Bykadorova~Telles$^{p}$,
J. J. Cherwinka$^{n}$,
T. Classen$^{e}$,
K. Commeford$^{b}$,
A. Conant$^{c}$,
D. Davee$^{o}$,
G. Deichert$^{h}$,
M. V. Diwan$^{a}$,
M. J. Dolinski$^{b}$,
A. Erickson$^{c}$,
B. T. Foust$^{p}$,
J. K. Gaison$^{p}$,
A. Galindo-Uribarri$^{i,k}$,
K. Gilje$^{d}$,
B. Hackett$^{i,k}$,
K. Han$^{p}$,
S. Hans$^{a}$,
A. Hansell$^{j}$,
K. M. Heeger$^{p}$,
B. Heffron$^{i,k}$,
J. Insler$^{b}$,
D. E. Jaffe$^{a}$,
D. Jones$^{j}$,
O. Kyzylova$^{b}$,
C. E. Lane$^{b}$,
T. J. Langford$^{p}$,
J. LaRosa$^{g}$,
B. R. Littlejohn$^{d}$,
F. Lopez$^{p}$,
D. A. Martinez~Caicedo$^{d}$,
J. Matta$^{i}$,
R. D. McKeown$^{o}$,
M. P. Mendenhall$^{e}$,
J. Minock$^{b}$,
P. E. Mueller$^{i}$,
H. P. Mumm$^{g}$,
J. Napolitano$^{j}$,
R. Neilson$^{b}$,
J. A. Nikkel$^{p}$,
D. Norcini$^{p}$,
S. Nour$^{g}$,
D. A. Pushin$^{l}$,
X. Qian$^{a}$,
E. Romero-Romero$^{i,k}$,
R. Rosero$^{a}$,
P. T. Surukuchi$^{d}$,
C. Trinh$^{b}$,
M. A. Tyra$^{g}$,
J. M. Wagner$^{b}$,
C. White$^{d}$,
J. Wilhelmi$^{j}$,
T. Wise$^{p}$,
M. Yeh$^{a}$,
Y.-R. Yen$^{b}$,
A. Zhang$^{a}$,
C. Zhang$^{a}$,
X. Zhang$^{d}$ \\
\llap{$^{a}$}Brookhaven National Laboratory, Upton, New York 11973, USA \\
\llap{$^{b}$}Department of Physics, Drexel University, Philadelphia, PA, USA \\
\llap{$^{c}$}George W. Woodruff School of Mechanical Engineering, Georgia Institute of Technology, Atlanta, GA USA \\
\llap{$^{d}$}Department of Physics, Illinois Institute of Technology, Chicago, IL, USA \\
\llap{$^{e}$}Nuclear and Chemical Sciences Division, Lawrence Livermore National Laboratory, Livermore, CA, USA \\
\llap{$^{f}$}Department of Physics, Le Moyne College, Syracuse, NY, USA \\
\llap{$^{g}$}National Institute of Standards and Technology, Gaithersburg, MD, USA \\
\llap{$^{h}$}High Flux Isotope Reactor, Oak Ridge National Laboratory, Oak Ridge, TN, USA \\
\llap{$^{i}$}Physics Division, Oak Ridge National Laboratory, Oak Ridge, TN, USA \\
\llap{$^{j}$}Department of Physics, Temple University, Philadelphia, PA, USA \\
\llap{$^{k}$}Department of Physics and Astronomy, University of Tennessee, Knoxville, TN, USA \\
\llap{$^{l}$}Institute for Quantum Computing and Department of Physics and Astronomy, University of Waterloo, Waterloo, ON, Canada \\
\llap{$^{m}$}Department of Physics, University of Wisconsin, Madison, Madison, WI, USA \\
\llap{$^{n}$}Physical Sciences Laboratory, University of Wisconsin, Madison, Madison, WI, USA \\
\llap{$^{o}$}Department of Physics, College of William and Mary, Williamsburg, VA, USA \\
\llap{$^{p}$}Wright Laboratory, Department of Physics, Yale University, New Haven, CT, USA \\
E-mail: \email{prospect.collaboration@gmail.com}
}

\abstract{
This paper describes the design and performance of a 50\,liter, two-segment $^{6}$Li-loaded liquid scintillator detector that was
designed and operated as prototype for the PROSPECT (Precision Reactor
Oscillation and Spectrum) Experiment. The two-segment detector was
constructed according to the design specifications of the
experiment. It features low-mass optical separators, an integrated source and optical calibration system, and materials that are
compatible with the $^{6}$Li-doped scintillator developed by PROSPECT. We
demonstrate a high light collection of 850$\pm$20\,PE/MeV, an energy
resolution of  $\sigma$ = 4.0$\pm$0.2\% at 1\,MeV, and
efficient pulse-shape discrimination of low $dE/dx$ (electronic recoil) and high
$dE/dx$ (nuclear recoil) energy depositions. An effective
scintillation attenuation
length of 85$\pm$3\,cm is measured in each segment. The 0.1\% by mass
concentration of $^{6}$Li in the scintillator results in a
 measured neutron capture time of $\tau$ = 42.8$\pm$0.2\,$\mathrm{\mu
   s}$. The long-term stability of the scintillator is also
 discussed. The detector response meets the criteria necessary for achieving the PROSPECT physics goals and demonstrates features that
 may find application in fast neutron detection.
}


\begin{document}



\section{Introduction} 
\label{sec:introduction}
Liquid scintillator detectors have been an important technology in neutrino physics since the first detection of (anti)neutrinos by Cowan and Reines at the Savannah River reactor in 1959\,\cite{Reines:1960}. 
Composed mostly of hydrocarbons, organic scintillators feature a high density of protons to facilitate inverse beta decay (IBD) interactions $\mathrm{\overline{\nu}_{e} + p \rightarrow e^{+} + n}$, efficiently moderate neutrons, and allow for cost-effective volume scaling. 
Recent advances in non-toxic and non-flammable scintillators with pulse-shape discriminating (PSD) capabilities, a phenomenon first observed by Brooks\,\cite{Brooks:1959}, alters the emission of scintillation light with a time structure that is heavily dependent on the energy deposition as a function of distance traveled by the ionizing particle. 
When combined with an appropriate neutron capture agent, PSD provides excellent discrimination between many backgrounds and neutron captures\,\cite{Bowden:2012um}. 
Moreover when the capture is correlated with a prompt positron-like event, this technique offers powerful selection of IBD interactions.
Although dependent on the light transport properties of materials used, PSD performance is strongly coupled to the detected photon statistics\,\cite{p20paper}. 
Moreover, scintillators with high light yield and detectors with correspondingly high collection efficiency and resolution are crucial for new precision neutrino experiments where a detailed measurement of the energy spectrum shape is desired.

Homogeneous ton-scale detectors are unable to fully capture $\gamma$-ray cascades induced by neutron captures on Gd (see, Ref.\,\cite{Classen:2014uaa}) , a common dopant utilized by large reactor experiments such as Daya Bay\,\cite{Beriguete:2014gua}, Double Chooz\,\cite{Aberle:2011ar}, and RENO\,\cite{Park:2013nsa}.
$^{6}$Li-loaded liquid scintillators ($^{6}$LiLS) are an ideal detection medium for compact detectors because the energetic, heavy ion products from neutron captures on $^{6}$Li, $\mathrm{n + ^{6}Li \rightarrow \alpha + t + }$ 4.78\,MeV, have a range of $\sim$100\,$\mathrm{\mu}$m. 
This reduces the amount of spatial variation in the neutron capture tagging efficiency.
Small 100\,ml-scale $^{6}$LiLS PSD detectors have recently been developed within the neutron physics community to determine neutron detection efficiencies\,\cite{Fisher:2011hm, Bass:2012ur, Zaitseva:2013}.
Larger segmented $^{6}$LiLS detectors are not widely used within neutrino physics, with the notable exception of the Bugey-3 reactor neutrino experiment\,\cite{Abbes:1995nc}. 
Thus, there is a need to understand the feasibility of segmented neutrino detectors based on $^{6}$LiLS. 

The PROSPECT experiment uses a segmented $^{6}$LiLS detector volume as both the target and detection medium to observe reactor antineutrinos from the High Flux Isotope Reactor (HFIR) at Oak Ridge National Laboratory (ORNL)\,\cite{Ashenfelter:2015uxt}. 
PROSPECT will address the reactor antineutrino flux anomaly\,\cite{Mueller:2011nm, Huber:2011wv} and the reactor spectrum anomaly\,\cite{Abe:2014bwa, Seo:2016uom, An:2015nua} by searching for eV-scale sterile neutrinos and precisely measuring the $^{235}$U antineutrino energy spectrum. 
The 4-ton $^{6}$LiLS detector is optically divided into an 11$\times$14 array read out on both ends by photomultiplier tubes (PMTs). 
As PROSPECT will operate with minimal overburden, a targeted shielding package consisting of layers of hydrogenous material, borated polyethylene, and lead has been designed.
In addition, active background suppression from PSD and segmentation work to reject cosmogenic and reactor-related backgrounds.
To maximize background rejection and meet the physics goals of PROSPECT, excellent light collection and energy resolution are required.

In operation at Yale Wright Laboratory, PROSPECT-50 is a two-segment, 50\,liter prototype.  
Each segment measures 117.6$\times$14.5$\times$14.5\,cm$^{3}$ and contains an active $^{6}$LiLS volume of 25\,liters. 
The detector was built with production parts to model the PROSPECT low-mass optical lattice and inner-detector calibration access. 
Materials were chosen to be compatible with the $^{6}$LiLS. 
We report the light collection, energy resolution, and PSD capabilities of the detector. 
The effective attenuation length of the $^{6}$LiLS, position reconstruction, and neutron capture time are also studied. 
A discussion of the long-term stability of the $^{6}$LiLS is provided. 
The PROSPECT-50 prototype demonstrates the feasibility and performance of a large-scale $^{6}$LiLS segmented detector.


\section{$^{6}$Li-loaded liquid scintillator}
\label{sec:scintillator}
Scintillation light in organic scintillators is produced by a series of molecular transitions induced by ionized valence electrons. 
Depending on the particular scintillator, this fluorescence can be composed of multiple radiative lifetimes between excited singlet and triplet states\,\cite{Birks:1964zz, Brooks:1979st}. 
The ratio of the prompt to delayed light is highly correlated to the stopping power $dE/dx$ of the primary ionizing particle, which can be translated via a detection system into waveforms with different charge integrals in the signal tail. 
Low $dE/dx$ (electronic recoil) and high $dE/dx$ (nuclear recoil) energy depositions can then be distinguished via the simple metric:

\begin{equation}
PSD = \frac{Q_{tail}}{Q_{full}},
\label{eq:PSD}
\end{equation}

\noindent where $Q_{full}$ is the integrated charge in the full waveform and $Q_{tail}$ the integrated charge in the tail of the waveform defined to optimize the discrimination between ionizing particle types. 
PSD capabilities of a scintillator detector can be improved with an increase in detected photons, improving the counting statistics. 
However, in detectors where geometry plays a significant role in the light collection mechanism, the number of scatters (due to reflective wall material) can also affect the PSD performance, as studied in Ref.\,\cite{p20paper}. 
Neutrino detectors that require event localization and uniform IBD identification efficiency can be improved with the addition of a $^{6}$Li-dopant to the scintillator as discussed in Section\,\ref{sec:introduction}. 

\begin{figure}[h]
\begin{center}  
\includegraphics[width=.65\textwidth]{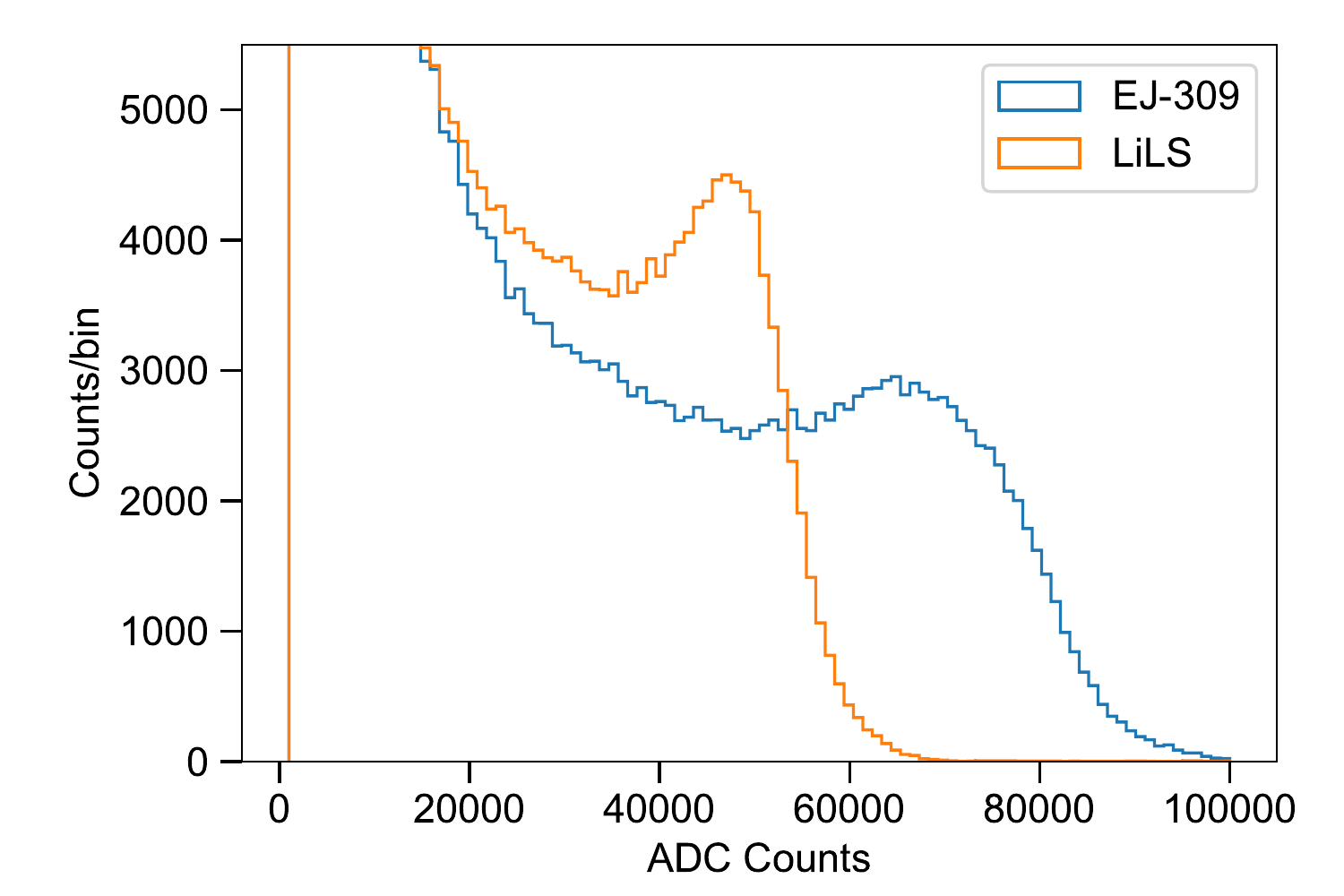} 
\caption{Light yield measurements of PROSPECT $^{6}$LiLS and EJ-309 as a reference using a 20\,ml vial test stand and a $\mathrm{^{137}Cs}$ source.}
\label{fig:nLi_LY}
\end{center}
\end{figure}

Previous $^{6}$Li-loaded scintillators used for IBD detection of reactor antineutrinos were pseudocumene based\,\cite{Ait-Boubker:1989}. 
However, these organic scintillators are toxic with low flash points and undesirable from a health and safety perspective for deployment near nuclear reactors. 
Through an extensive R\&D program, PROSPECT has developed a non-toxic, non-flammable $^{6}$LiLS based on EJ-309 from Eljen Technology\,\cite{disclaimer, ej309}. 
A surfactant is added to allow the loading of an aqueous 95\%-enriched $^6$LiCl solution in a dynamically stable microemulsion. 
The final $^{6}$Li concentration is 0.1\% by mass. 
A detailed study of microemulsion stability in $^{6}$LiLS can be found in Ref~\cite{Bergeron2017}. 
This process reduces the light yield of the EJ-309 by $\sim$30\% to $8200\pm$200\,photons/MeV, as shown in Figure\,\ref{fig:nLi_LY}.
Table\,\ref{table:ej309} lists the properties relevant for particle detection.

\begin{table}[h]
\begin{center}
\begin{tabular}{l | r | r}
\hline
& EJ-309 & PROSPECT $^{6}$LiLS \\
\hline
C:H:O & 90.6\%:9.4\%:0\%& 84.14\%:9.52\%:6.34\% \\
Density (g/ml) & 0.959 &0.979$\pm$0.001 \\
Maximum emission (nm) & 424 & 424\\
Light yield (photons/MeV) & 11,500\footnotemark & $8200\pm200$\\
\hline
\end{tabular}
\caption{Properties of EJ-309 scintillator given by Eljen Technology\,\cite{ej309} and PROSPECT $^{6}$LiLS, as measured by the collaboration. }
\label{table:ej309}
\end{center}
\end{table}
 \footnotetext{A newer version of EJ-309 has become available with a light yield of 12,300\,photons/MeV, which is currently quoted on the Eljen Technology website\,\cite{ej309}.}

The light yield and optical properties of scintillating organic solvents can be negatively affected by chemical interactions with the surrounding environment, such as detector materials or air near the detector liquid surface\,\cite{PhysRev.101.998, 1674-1137-34-5-011}.  
Through extensive compatibility testing, appropriate detector materials and operating environments have been determined for the PROSPECT $^{6}$LiLS.
Acrylics, PTFE, FEP, PLA, Viton, PEEK, and selected acrylic cements did not degrade the optical properties of the scintillator or the materials themselves, and were exclusively used in all locations in contact with the $^{6}$LiLS. 
Cast acrylic, Viton, PEEK, PTFE, and FEP were found to maintain structural integrity over extended time periods in the presence of the PROSPECT $^{6}$LiLS. 
At relatively low stress levels, extruded acrylic and PLA also maintained structural integrity.
In PROSPECT vial tests, a nitrogen gas cover blanket was shown to reduce oxygen quenching and improve PSD, in agreement with the literature\,\cite{OKeeffe:2011dex,Li:2011a,HuaLin:2009sg}. 
These results have informed the design of the PROSPECT detector and PROSPECT-50.




\section{PROSPECT-50 detector design}
\label{sec:detector}

\begin{figure}[h]
\begin{center}
\includegraphics[width=0.95\textwidth]{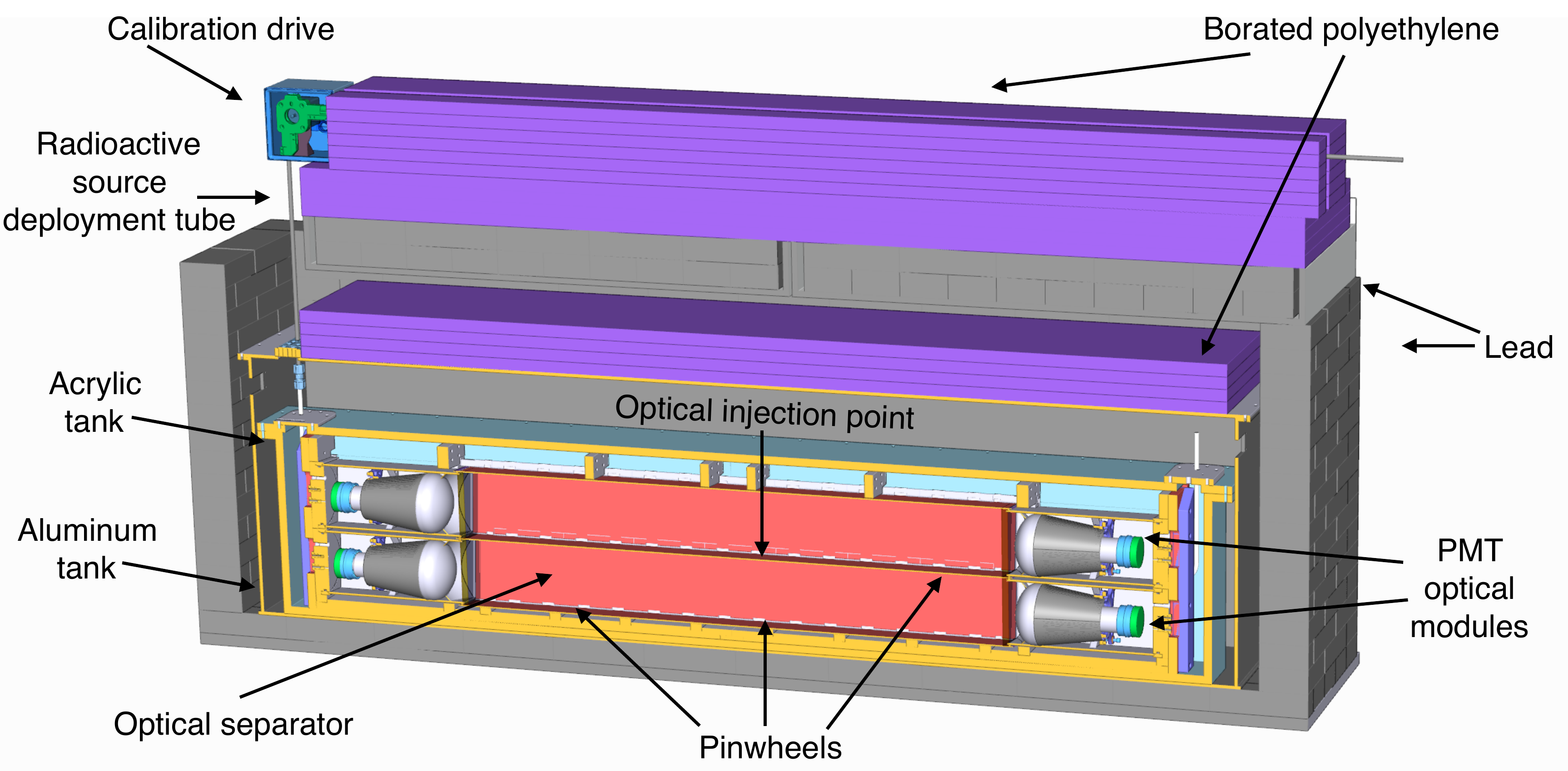}
\caption{The PROSPECT-50 detector design. The detector has two 117.6$\times$14.5$\times$14.5\,cm$^{3}$ optical segments (red), each viewed by two 12.7\,cm-diameter photomultiplier tubes (white/gray) contained in an acrylic enclosure filled with mineral oil for separation from the scintillator. Borated polyethylene (green/purple) and lead (dark gray) surround the detector.}
\label{fig:P50cad}
\end{center}
\end{figure}

PROSPECT-50 consists of two full-scale PROSPECT $^{6}$LiLS optical segments supported in an acrylic vessel inside of an aluminum secondary containment tank shielded by lead and borated polyethylene, as illustrated in Figure\,\ref{fig:P50cad}. Along with characterizing the scintillator performance, the detector has served as a testbed for the fabrication and operation of various subsystems that have been implemented in the PROSPECT detector, which are described below. 


\subsection{Optical modules}
PROSPECT uses a total of 308, 12.7\,cm (5\,in)-diameter hemispherical Hamamatsu R6594\,\cite{R6594} and Electron Tubes Ltd 9372\,\cite{9372} photomultipler tubes.
We designate these as HPK and ETL PMTs, respectively, in the remainder of this article. 
Both PMT types are used in PROSPECT-50 to provide a relative characterization of their performance. 
The PMTs use tapered voltage dividers to preserve linearity over a wide dynamic range.

\begin{figure}[h]
\begin{center}
\includegraphics[width=0.60\textwidth]{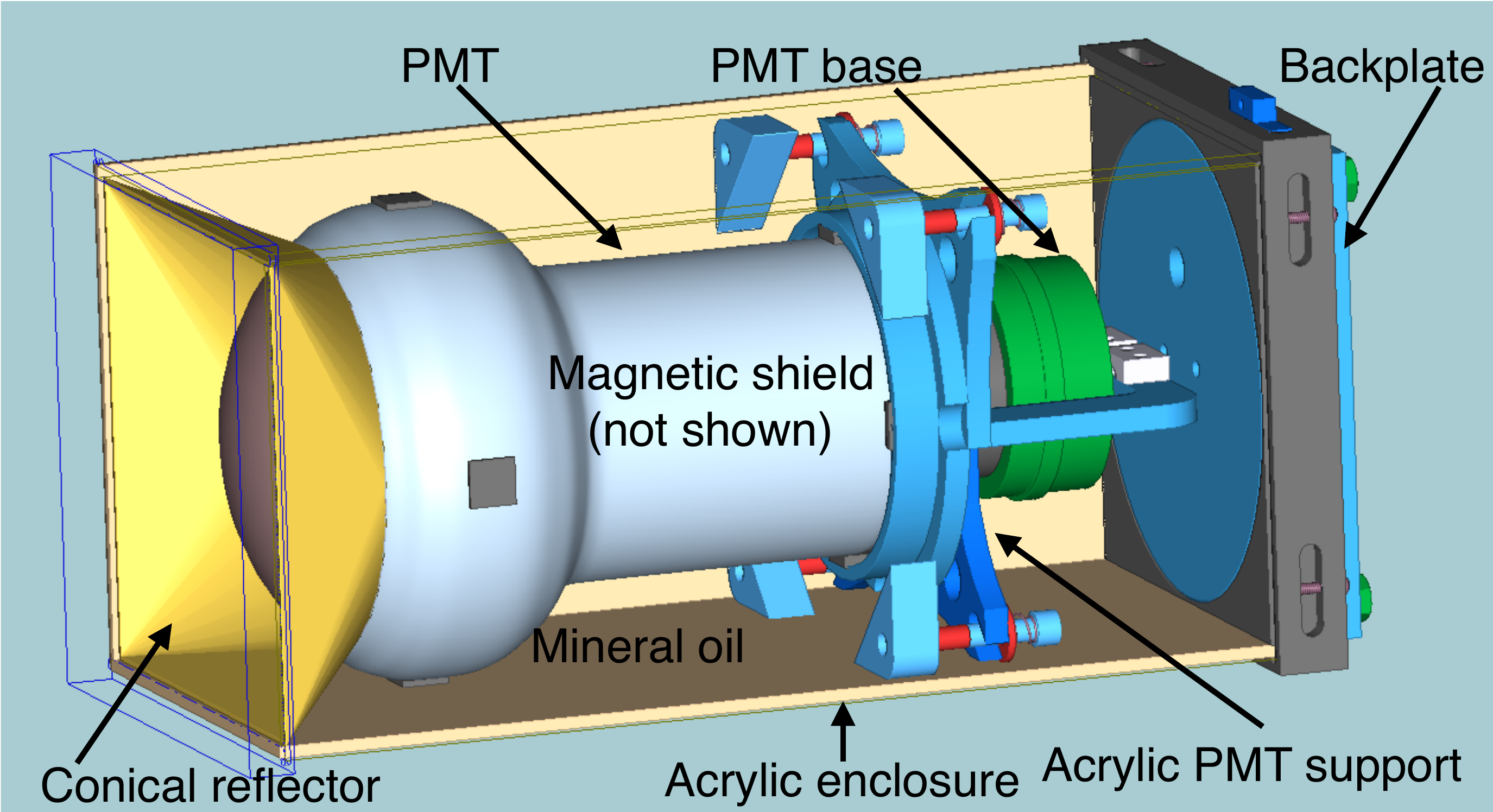}
\includegraphics[width=0.25\textwidth]{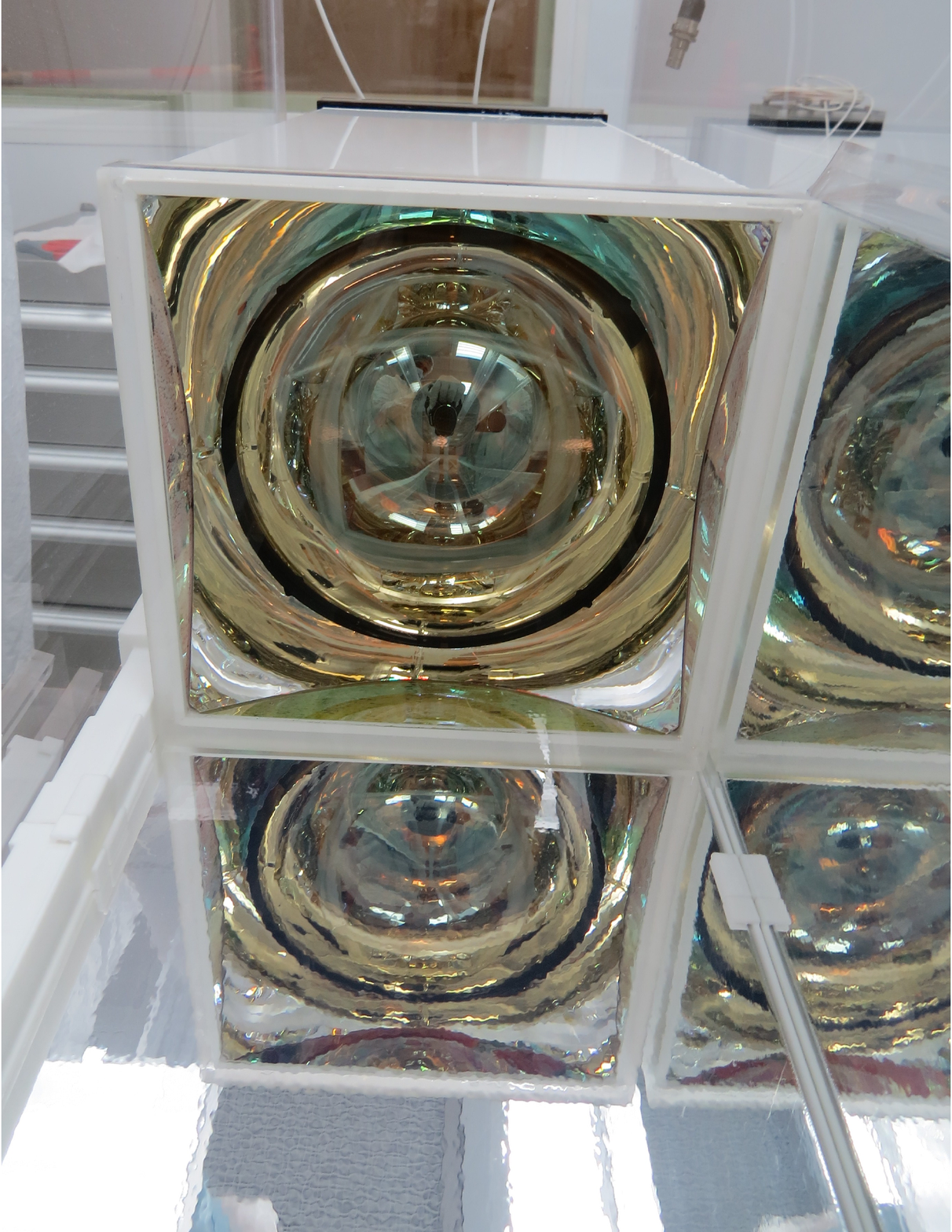}
\caption{\textbf{(a)} CAD rendering of a PROSPECT optical module. \textbf{(b)} Photograph of the front of an optical module in the PROSPECT-50 detector.}
\label{fig:Housing}
\end{center}
\end{figure}

Each PMT is housed in an acrylic enclosure filled with mineral oil to provide optical coupling to the scintillator volume, referred to as PROSPECT optical modules (POMs). 
As shown in Figure\,\ref{fig:Housing}, the assembled POMs are positioned directly in the liquid volume, and the exteriors are constructed of $^{6}$LiLS-compatible materials. 
The walls of the POMs are made from opaque white acrylic to minimize crosstalk between adjacent segments, while the front windows are made from UV-transmitting acrylic to maximize the transmission of the scintillation light. 
The backplate of the POM is made liquid-tight via Viton O-rings. 
PEEK plugs seal the penetrations needed for exit of the PTFE-coated signal and high voltage cables. 
The interior of each POM is outfitted with an acrylic frame to support the PMT with a Hitachi Finemet magnetic shield\,\cite{DeVore:2013xma} (not shown in Figure\,\ref{fig:Housing}).
A truncated conical reflector, laser-cut from thin acrylic sheet and then coated with 3M DF2000MA\,\cite{df2000ma} specular reflector film, is used to concentrate incident light onto the PMT photocathode. 
Slightly different shapes are employed for the two PMT types to optimize light collection given the manufacturer-specified photocathode dimensions.


\subsection{Optical lattice}
As noted above, the single-volume $^{6}$Li-loaded liquid scintillator target in the PROSPECT detector is optically divided into 154 identical segments. 
The segmentation of the detector is achieved by 1.5\,mm thick optical separators designed to reflect scintillation light to the POMs with limited crosstalk. 
These separators are built of multiple layers of material. High-gloss carbon fiber sheet is used as a backbone and provides the required rigidity and flatness. 
Specular reflector film is laminated on either side of the carbon fiber sheet and enclosed in a thin heat-bonded FEP film package to ensure compatibility with the $^{6}$LiLS.

\begin{figure}[h]
\begin{center}
\includegraphics[width=0.45\textwidth]{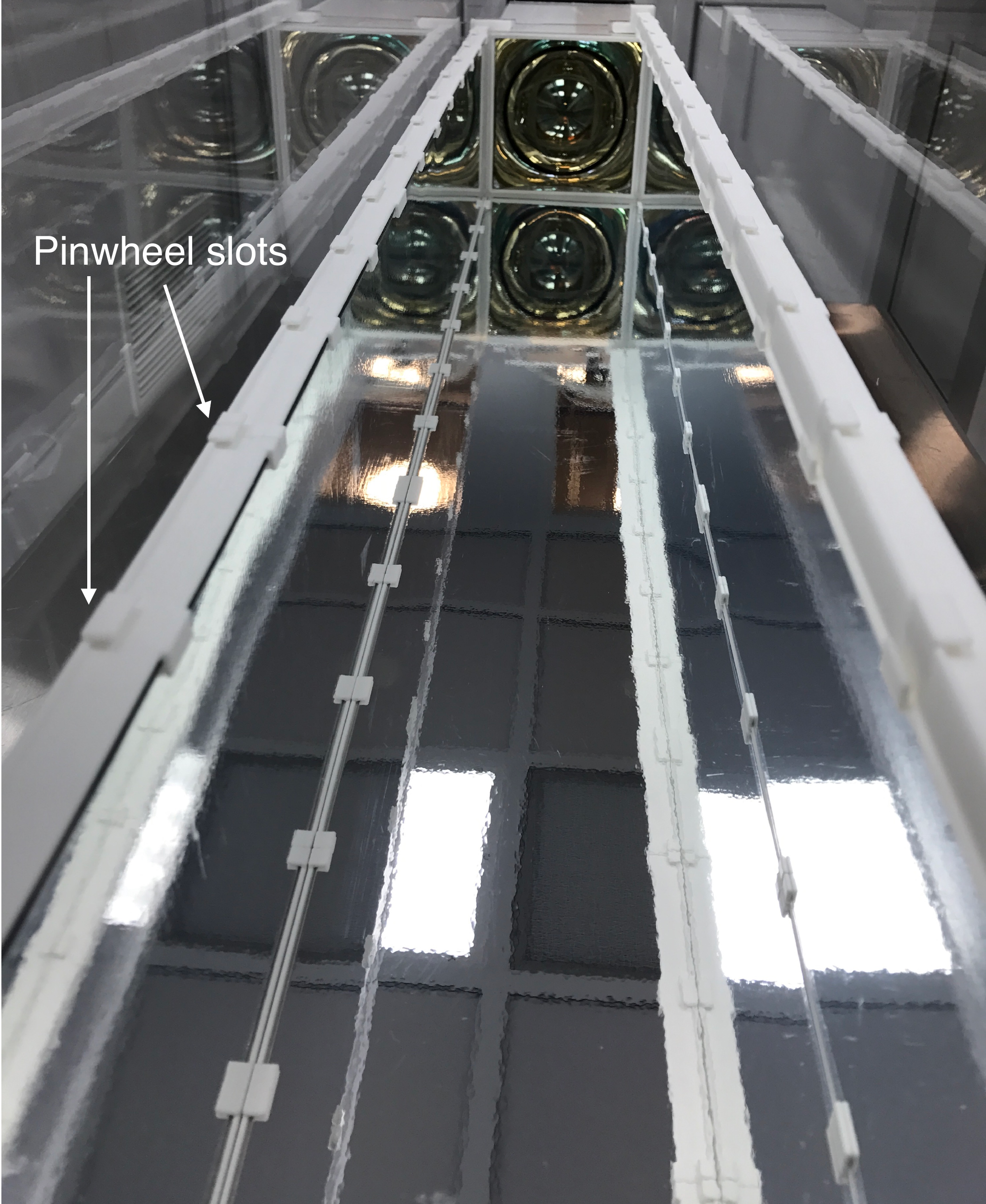}
\caption{Photograph of a partial optical lattice viewed by an installed optical module. The white pinwheels line the edges of the optical separators and join the adjacent separators (three shown). Some pinwheel slots that hold the separators in place are labeled. Reflections from the room light, ceiling, and optical module are clearly visible. }
\label{fig:optical_lattice}
\end{center}
\end{figure}

The optical separators are structurally supported at the ends by POMs and constrained along the length of the segment by filament-based 3D-printed white PLA plastic supports, called pinwheels. 
The pinwheels are strung on a PTFE tube or acrylic rod for alignment. 
Slots are provided at regular intervals on the pinwheels for the optical separators to be inserted. 
With over 60\% reflectivity and an overall PLA surface exposure of $<$1\% to the active reflective area, the pinwheels are designed to maintain the high optical transport performance of the segments. 
The pinwheels are hollow to allow for \textit{in-situ} radioactive source and optical calibrations between segments of the detector. 
The optical separators and pinwheels were designed to comprise $\sim$3\% of the total mass fraction of the PROSPECT detector active volume.


\subsection{Internal calibration}
\label{subsec:calibrationDesign}
A key feature of the PROSPECT design is the ability to deploy calibration sources throughout the detector, along the segment length and between each detector segment, without altering any characteristic of the optical volume.
This is achieved via PTFE tubes of 3/8\,inch outer diameter and 1/4\,inch inner diameter extending along the detector length inside the pinwheel supports. 
In PROSPECT-50, radioactive source deployment tubes have been installed between the two segments, illustrated in Figure\,\ref{fig:P50cad}. 
A thin timing belt with a source capsule on one end is driven through the calibration tube by a stepper motor through a feedthrough on the exterior aluminum tank. 
Compact $\gamma$-ray sources (4\,mm diameter) are deployed into the detector for periodic energy calibrations and position scans to study detector response. 
Neutron calibration is performed outside of the detector volume in PROSPECT-50, although capsule-based neutron sources will be used in the PROSPECT detector. 

The PROSPECT-50 optical calibration system (OCS) is based on a CAEN SP5601 LED pulse driver which provides a 5\,ns wide, 405\,nm wavelength light pulse of variable intensity\,\cite{sp5601}.
Light is transported from the pulser to the midpoint of one of the supports between the two segments using an optical fiber encased in a PTFE sheath. 
In the center of the pinwheel is a diffusor sub-assembly designed to equally illuminate all adjacent segments. 
Calibration runs were performed triggering solely on the OCS to determine the gain of each PMT and measure the timing offsets between all four PMTs.
Additionally, the LED driver was triggered at 20\,Hz throughout operation to monitor the single photoelectron (PE) response of the PMTs and allow for correction of gain drifts.

\begin{figure}[h]
\begin{center}
\includegraphics[width=0.49\textwidth]{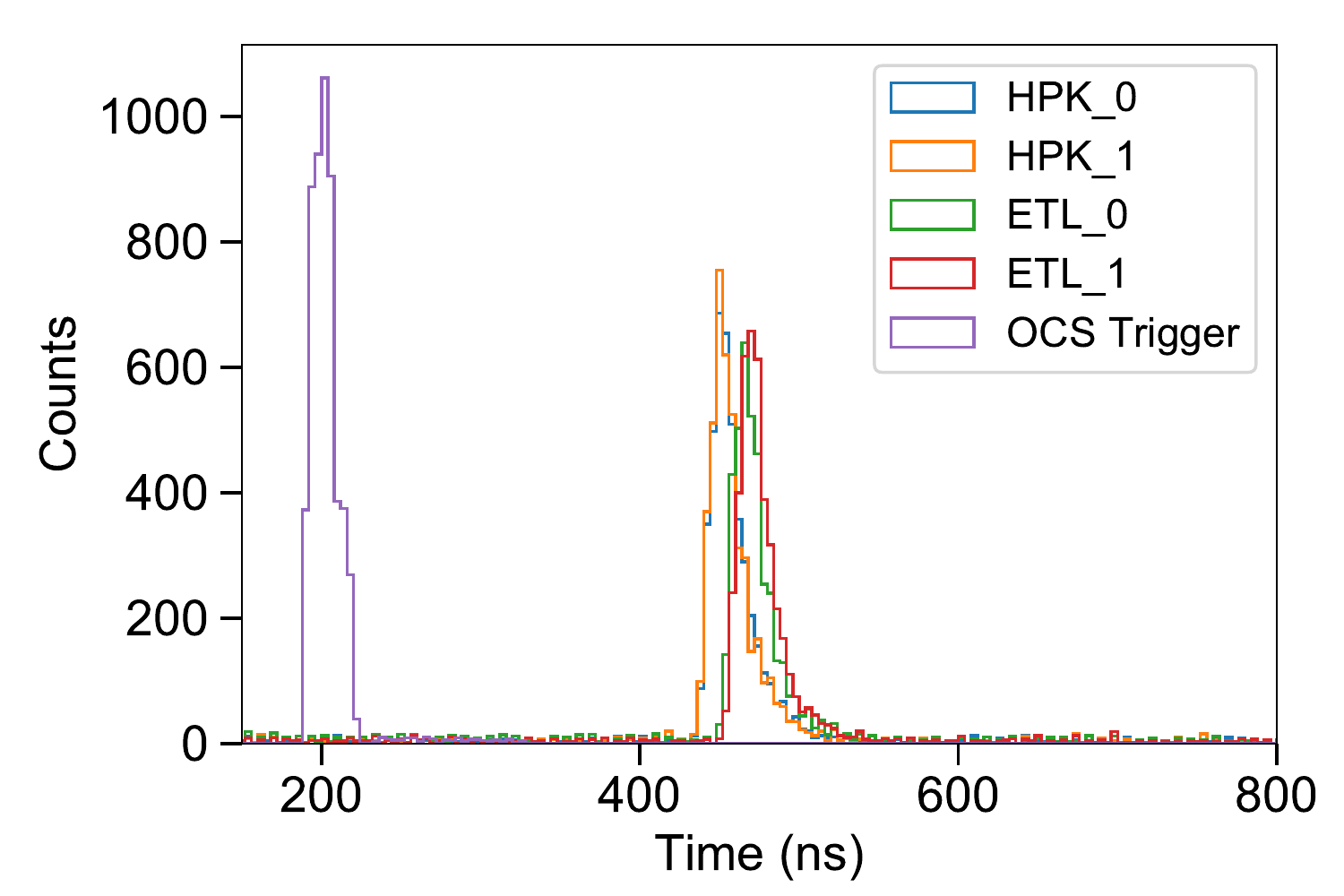}
\includegraphics[width=0.49\textwidth]{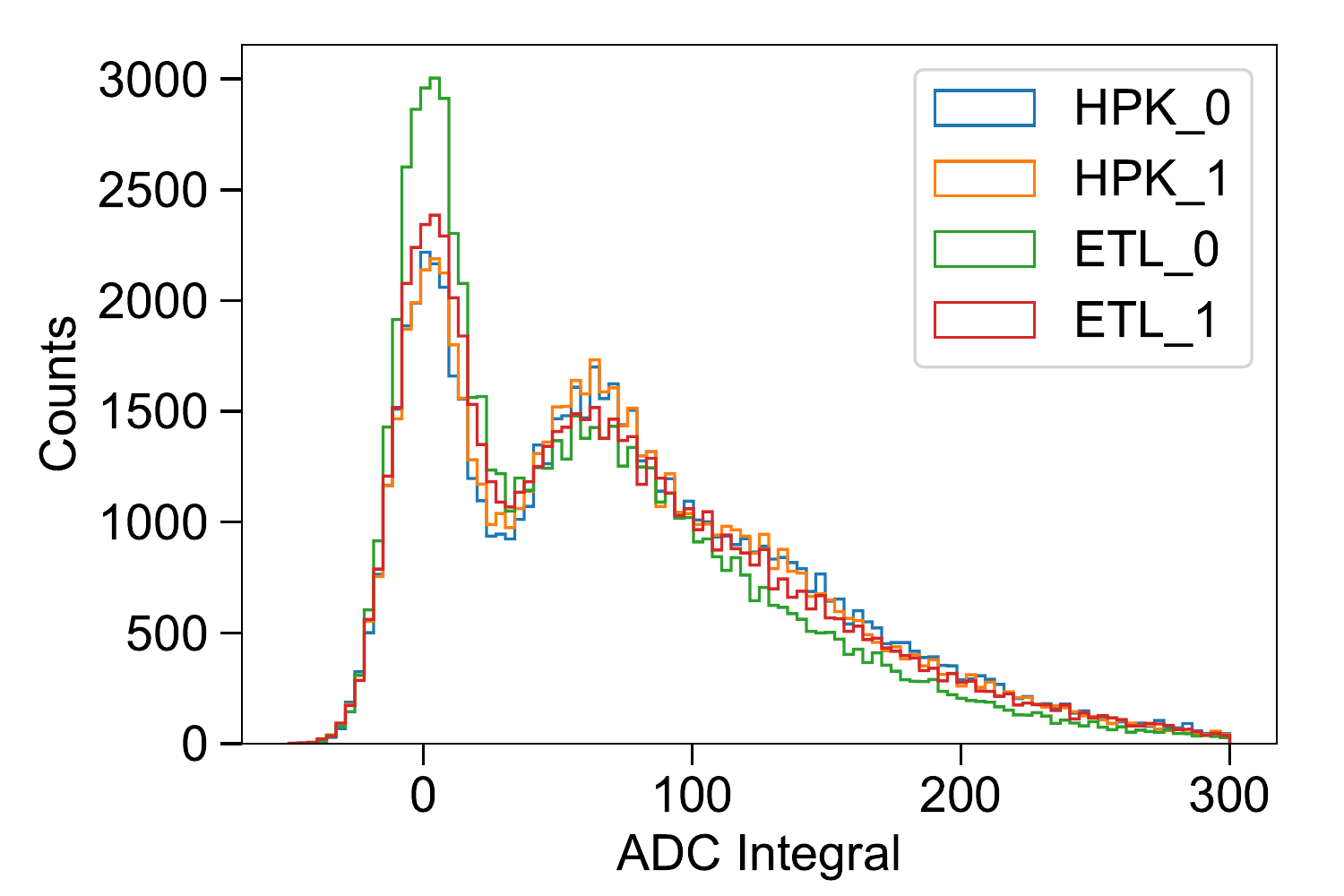}
\caption{ \textbf{(a)} Arrival time distributions of LED pulses for each PMT with the trigger pulse as a reference (labeled OCS Trigger). \textbf{(b)} Single and multi-photoelectron peaks for the Hamamatsu PMTs in the bottom segment (labeled HPK\_0 and HPK\_1) and the ETL PMTs in the top segment (labeled ETL\_0 and ETL\_1). A fit to each distribution yields a single photoelectron response of 60\,ADC/PE. }
\label{fig:spe}
\end{center}
\end{figure}

Figure\,\ref{fig:spe} demonstrates the timing distribution of the LED pulses. 
The ETL and HPK PMTs have different average transit times, leading to timing offsets between segments.
Small offsets from cable length differences and electronics delays are calibrated and accounted for in subsequent energy and position reconstructions. 
Distributions of recorded signal integrals from OCS-triggered events are shown for each PMT. 
The gain of each PMT was adjusted such that all single photoelectron peaks align at 60\,ADC/PE. 
This conversion is used to determine the number of detected photoelectrons in the analysis described in Section\,\ref{sec:response}. 

\subsection{Shielding configuration}
PROSPECT-50 is surrounded by a shielding package designed to reduce the two main sources of ambient backgrounds: local $\gamma$-ray activity and cosmogenic fast neutrons. 
To reduce the flux of ambient $\gamma$-rays from the environment, a 10\,cm thick layer of lead surrounds the entire detector package, as shown in Figure\,\ref{fig:P50cad}. 
Borated polyethylene is placed above the detector to shield downward-directed fast neutrons. 
The neutron shielding is split into two layers, 33\,cm above the top lead shield and 10\,cm between the lead and the detector. 
The inner neutron shielding reduces the flux of neutrons produced by spallation in the lead shield. 
This style of layered shielding material is based on the design of the shielding package for use by the PROSPECT detector at HFIR, established in Ref.\,\cite{Ashenfelter:2015tpm}. 

\subsection{Data acquisition and environmental monitoring}
Signals from the four PMTs are recorded by a CAEN V1725 waveform digitizer with 250\,MS/s sampling frequency and 14\,bit resolution\,\cite{caen_web_v1725}. 
The data acquisition system is triggered internally when both PMTs in a segment cross a set threshold. 
An acquisition window of 1.2\,$\mu$s is collected for each trigger and recorded for off-line analysis. 
An external trigger is produced by the OCS pulser and results in the digitization of all four PMT signals regardless of occupancy.

Environmental conditions are continuously monitored by temperature, pressure, and humidity sensors. 
A $\sim$20~liter expansion volume above the liquid surface underneath the acrylic vessel lid is maintained with nitrogen cover gas throughout the lifetime of detector to prevent oxygen ingress that could degrade the scintillator performance.

An alarm system produces alerts when any of the parameters depart from the normal operating ranges.


\section{Detector performance}
\label{sec:response}
Detector performance metrics were studied using radioactive sources positioned immediately outside of the detector ($\mathrm{^{252}Cf}$) and internally ($\mathrm{^{137}Cs}$) via the calibration system described in Section\,\ref{subsec:calibrationDesign}. 
Unless otherwise noted, the data below represent results from the segment with HPK PMTs, as the ETL segment was found to perform similarly. 
The data presented here were taken in the first months of detector operation.


\subsection{Attenuation and position reconstruction}
\begin{figure}[h!]
\begin{center}
\includegraphics[width=0.49\textwidth]{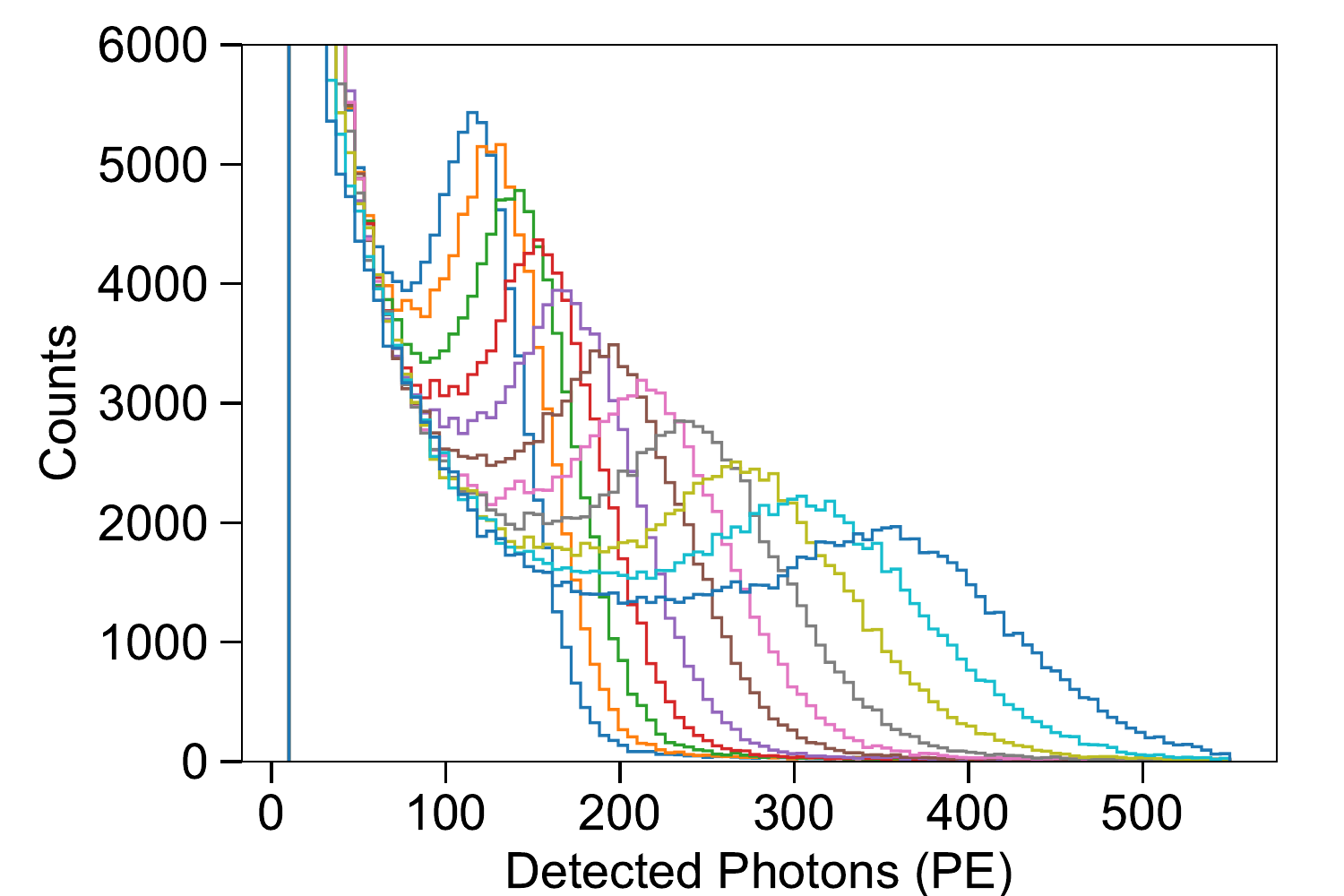} 
\includegraphics[width=0.49\textwidth]{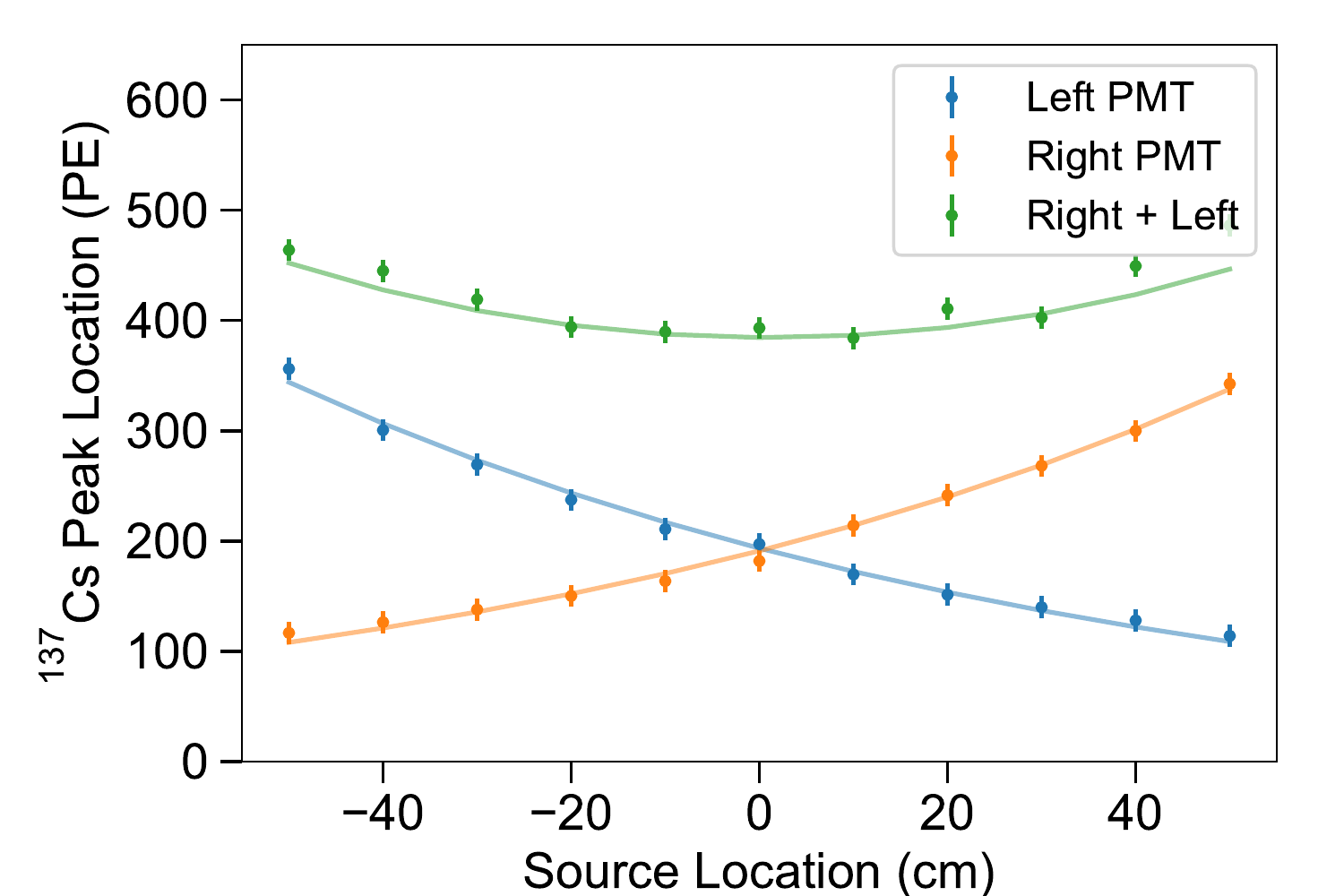} 
\caption{\textbf{(a)} Single-ended PMT $\mathrm{^{137}Cs}$ spectra in photoelectrons. The source was positioned in 10\,cm increments along the segment axis. \textbf{(b)} Average number of detected photoelectrons at the various positions with a fit attenuation length of 85$\pm$3\,cm.}
\label{fig:attenuation}
\end{center}
\end{figure}

The light collection efficiency varies along the segment length due to bulk attenuation of the $^{6}$LiLS, imperfect reflective surfaces, and other geometric effects. 
These combined losses will be referred to as the effective attenuation of the scintillator. 
To quantify this variation, the $\mathrm{^{137}Cs}$ $\gamma$-ray source was deployed in increments of 10\,cm along the segment (with 0\,cm indicating the center). 
Figure\,\ref{fig:attenuation} illustrates the collected photoelectron response for one PMT in the segment where the number of detected photoelectrons increases as the source approaches the PMT face. 
The maximum of the Compton spectrum extracted for individual PMTs (Right PMT, Left PMT) and with the charge from the two PMTS added (Right + Left). 
Higher light collection is observed at the segment ends with the minimum at the segment center. 
Fitting both single-ended PMT collection curves with simple exponentials, $Ae^{(-x/\lambda)}+C$, gives an effective attenuation length of $\mathrm{\lambda}$ = 85$\mathrm{\pm}$3\,cm. 

\begin{figure}[h!]
\begin{center}
\includegraphics[width=0.49\textwidth]{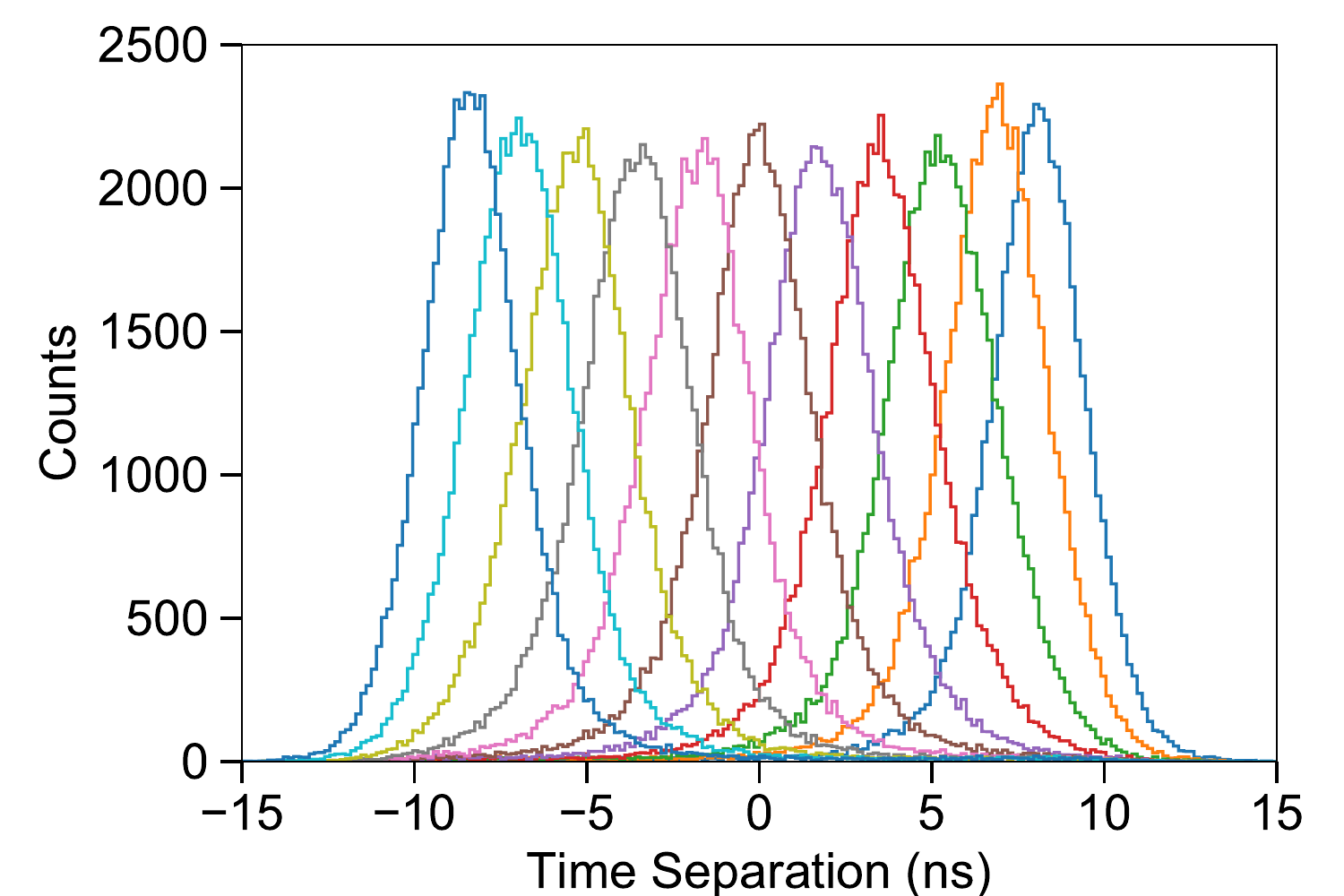}
\includegraphics[width=0.49\textwidth]{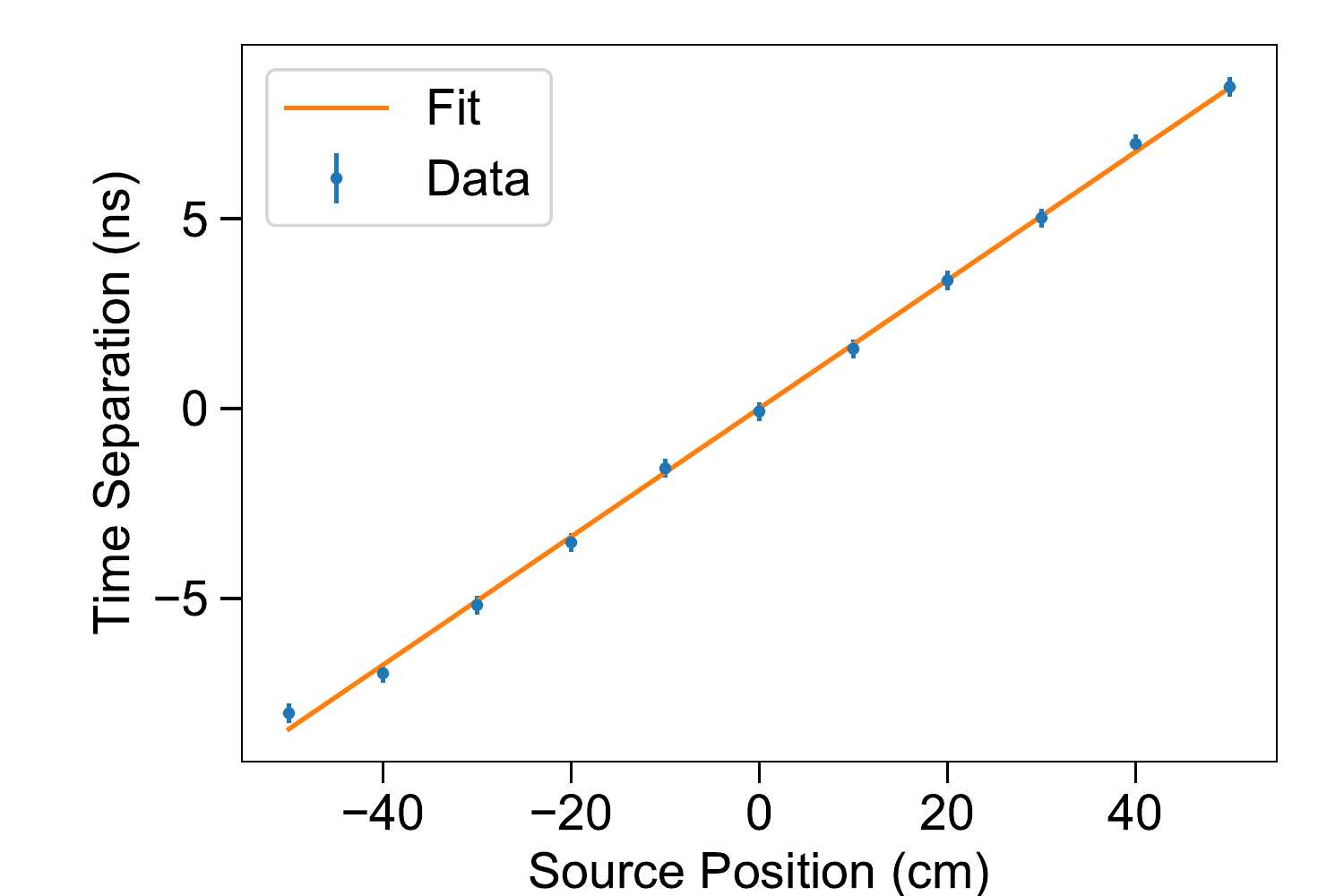}
\caption{\textbf{(a)} Time separation distributions of $\mathrm{^{137}Cs}$ events along the segment. \textbf{(b)} Comparison of deployed $^{137}$Cs position, with respect to the center of the segment, to the peak time separation.}
\label{fig:position_reconstruction}
\end{center}
\end{figure}

The non-uniformity in collection can be corrected by position-based reconstruction.
The difference between signal arrival times in the two PMTs on each segment provides a strong measure of the interaction position.
Figure\,\ref{fig:position_reconstruction} shows timing distributions for the $^{137}$Cs source deployed along the segment length. 
The width of each distribution is due to variation of interaction sites for the uncollimated $\gamma$-rays.
A linear reconstruction of the source position is observed solely using the timing metric. 
These measurements were used to develop a position reconstruction model based on a resolution-weighted average of the timing difference and log-ratio of charge for the two PMTs.




\subsection{Energy response}
\begin{figure}[h!]
\begin{center}
\includegraphics[width=0.49\textwidth]{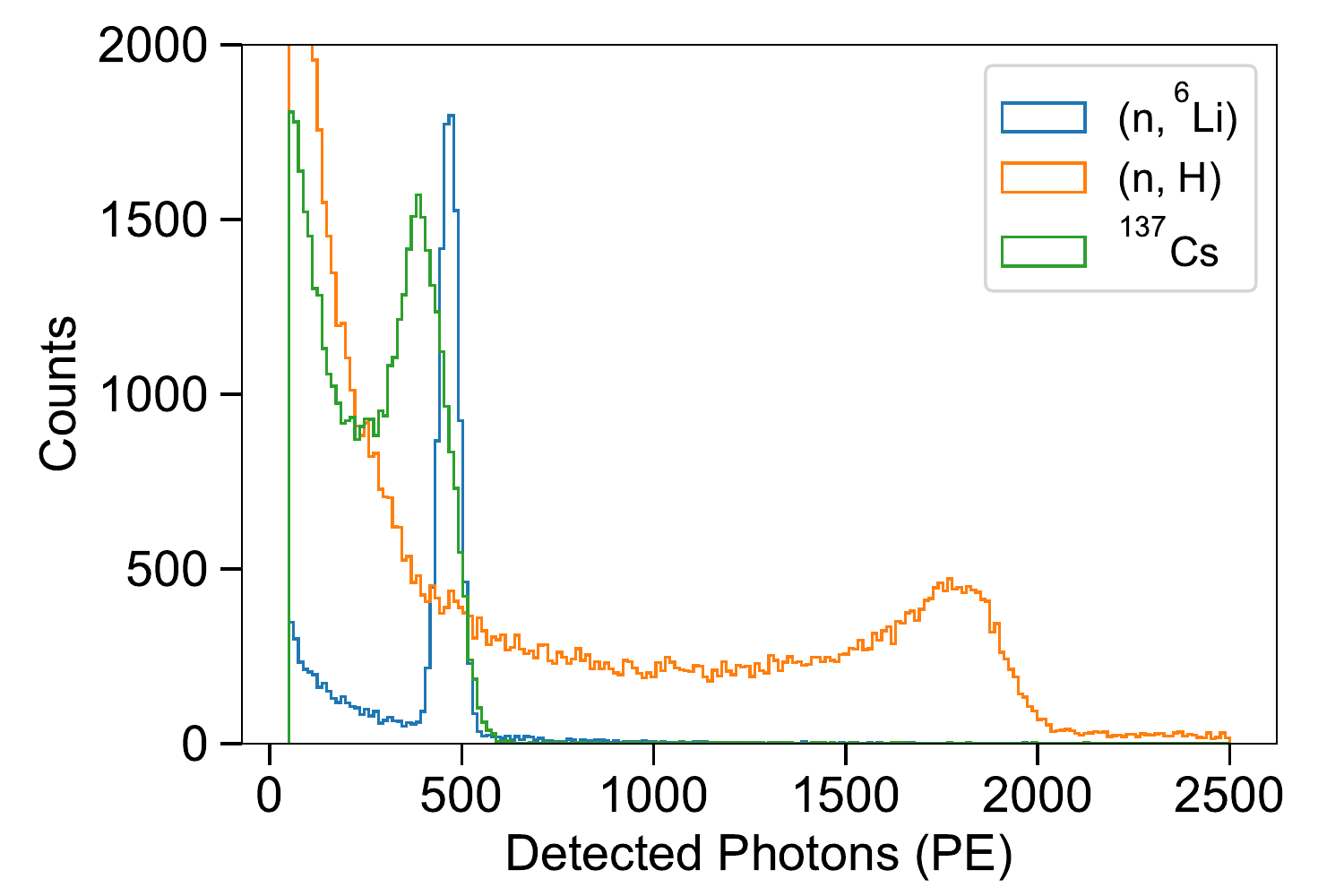} 
\includegraphics[width=0.49\textwidth]{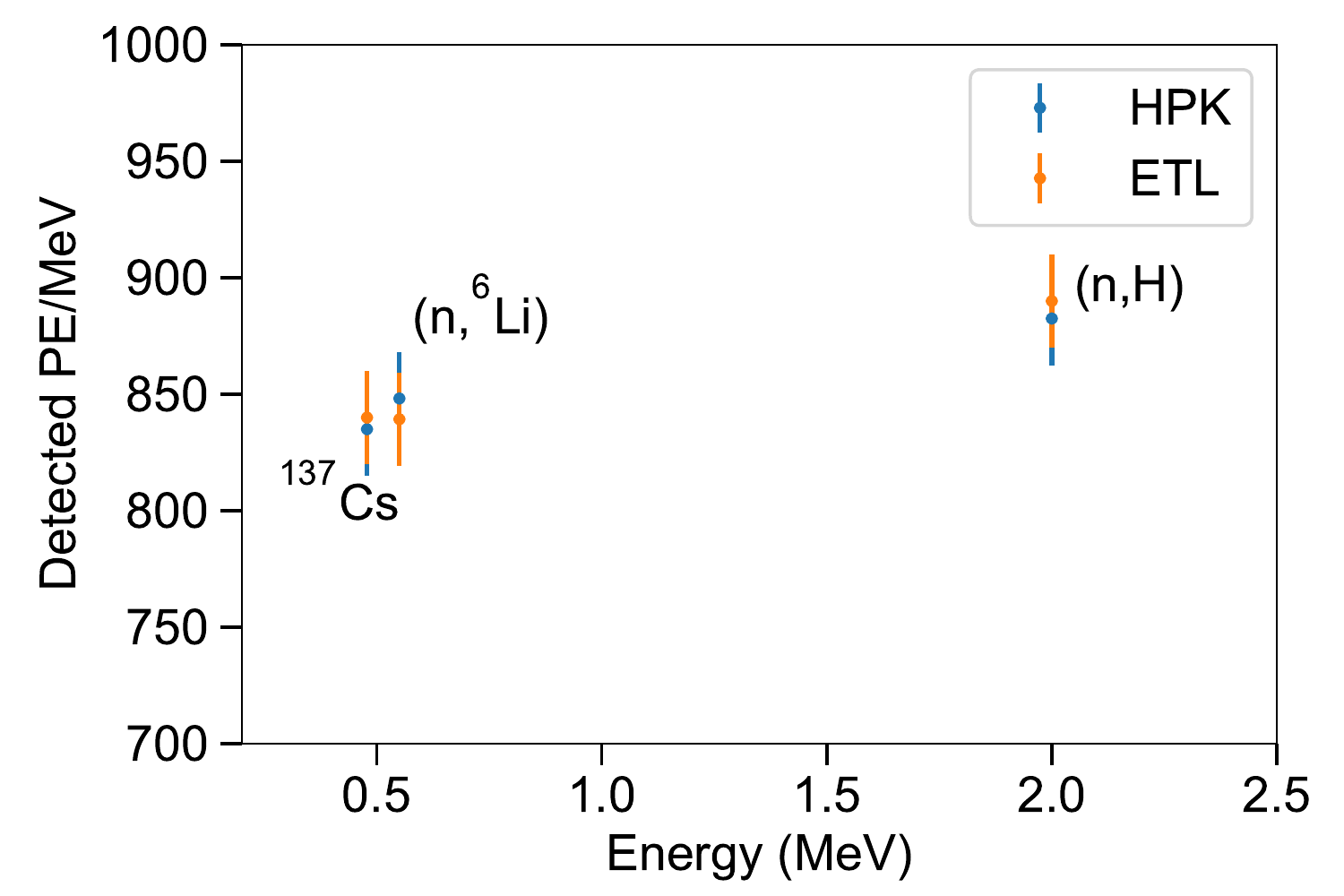} 
\caption{\textbf{(a)} $\mathrm{^{252}Cf}$ and $\mathrm{^{137}Cs}$
  energy spectra in units of photoelectrons highlighting the peak of
  (n,$^{6}$Li) and Compton edges of (n,H) and $\mathrm{^{137}Cs}$. \textbf{(b)} Light collection in terms of photoelectron/MeV for $\mathrm{^{252}Cf}$ and $\mathrm{^{137}Cs}$ features. Shown in (blue) are the results from the HPK segment and in (orange) results from the ETL segment.}
\label{fig:spectra}
\end{center}
\end{figure}

Scintillation from $\mathrm{\alpha}+t$ pairs due to neutron captures on lithium (n,$^{6}$Li) ($\mathrm{\sim}$0.55\,MeV electron equivalent energy), the $\gamma$-ray due to neutron captures on hydrogen (n,H) (2.2\,MeV), and the $\gamma$-ray from the decay of $\mathrm{^{137}Cs}$ (0.662\,MeV) are used to understand the $^{6}$LiLS energy response. 
A position-dependent light collection efficiency has been applied to the data based on the above measurements. 
The resulting spectra, in photoelectrons, from $\mathrm{^{252}Cf}$ and $\mathrm{^{137}Cs}$ selecting on the three event types are shown in Figure\,\ref{fig:spectra}.

\begin{figure}[h!]
\begin{center}
\includegraphics[width=0.49\textwidth]{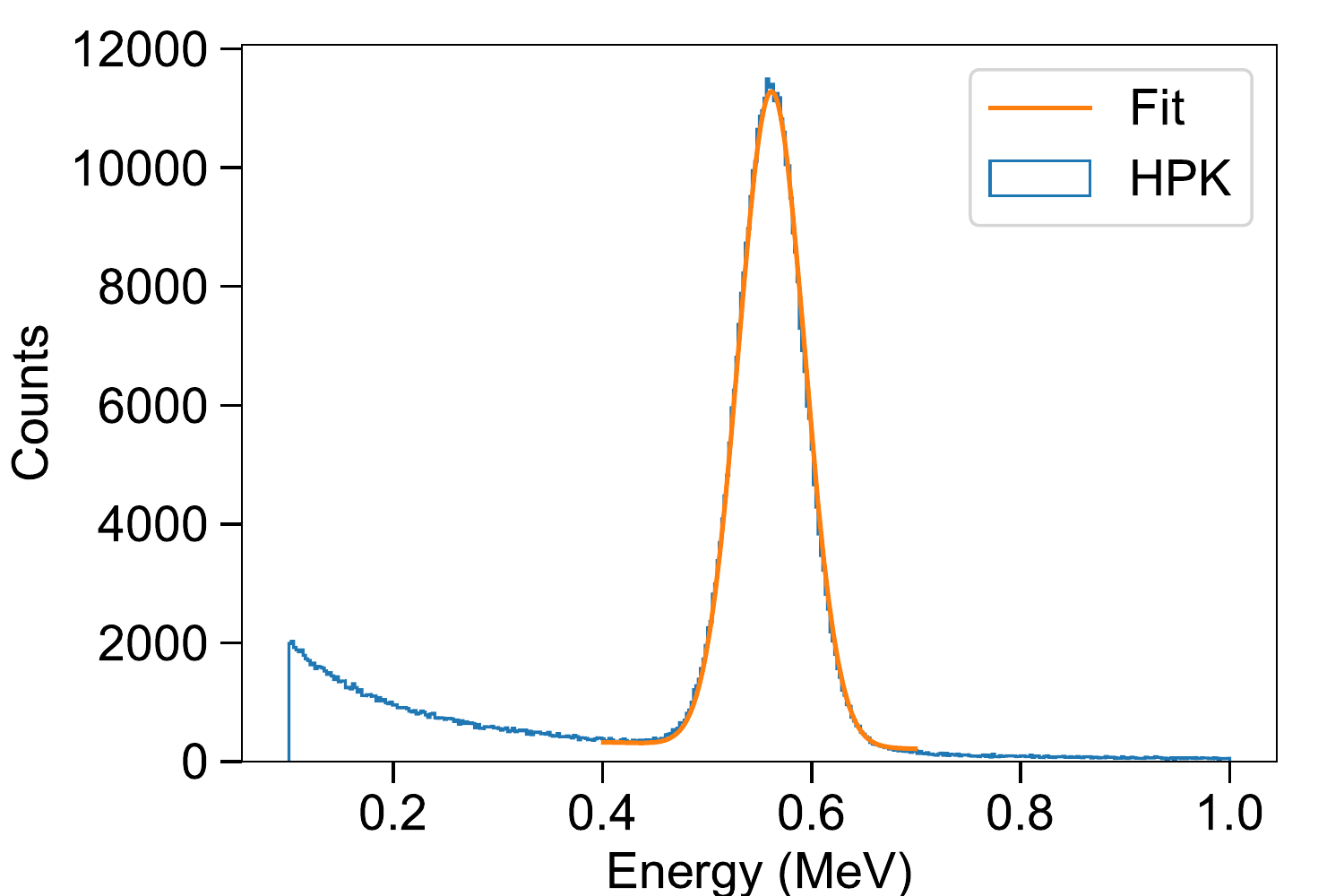}
\includegraphics[width=0.49\textwidth]{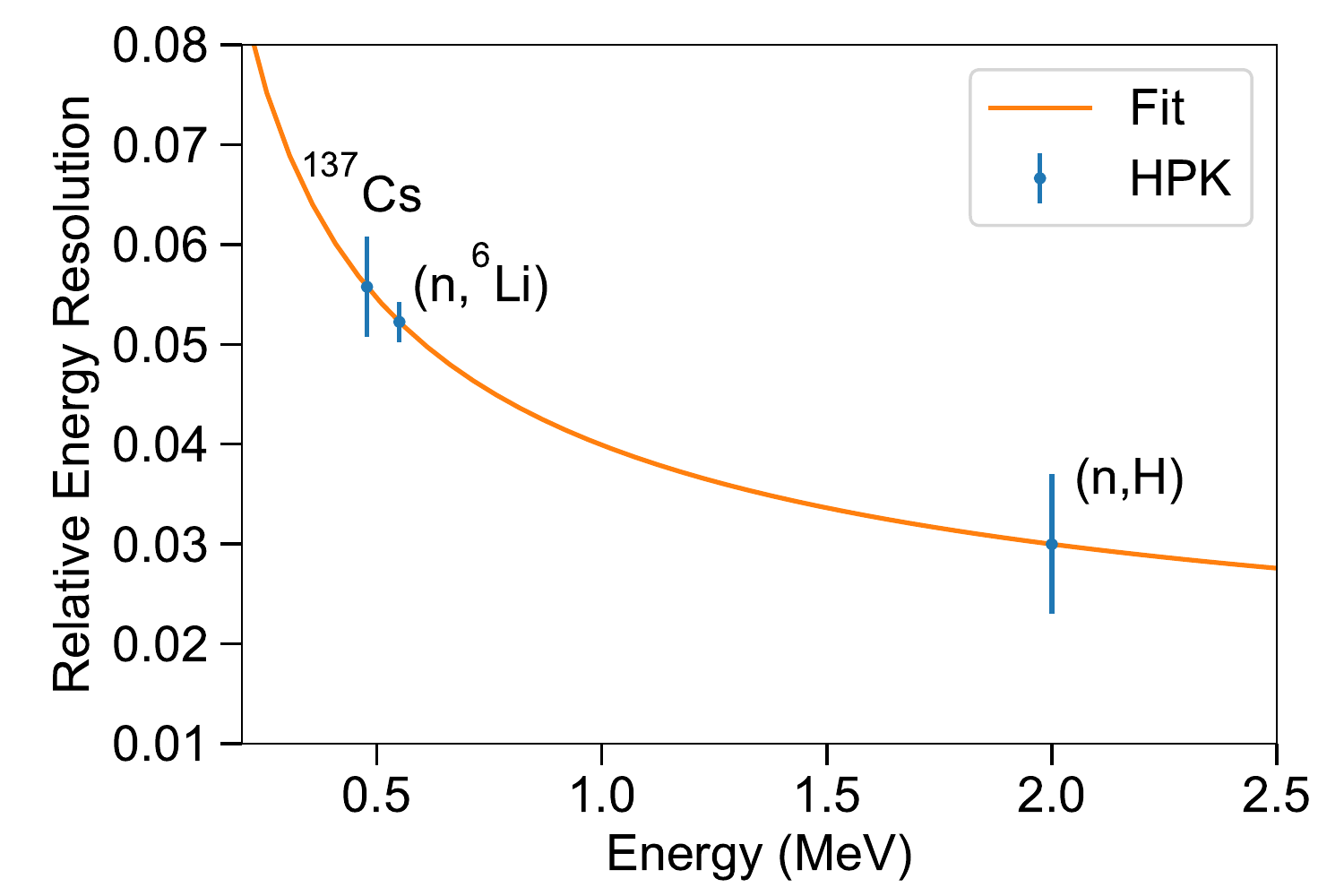}
\caption{\textbf{(a)} The (n,$^{6}$Li) peak fit with a Gaussian to extract an energy resolution. \textbf{(b)} Relative energy resolution as a function of nominal energy for (n,$^{6}$Li), (n,H), and $\mathrm{^{137}Cs}$ features. The fit contains a term for photon statistics and position-dependent collection effects. A resolution of  $\sigma$ = 4.0$\pm$0.2\% at 1\,MeV is measured.}
\label{fig:nLi_fit}
\end{center}
\end{figure}

The light collection efficiency for each event type was obtained by comparing the measured spectra to simulated response spectra (resolution smeared energy deposition spectra from Monte Carlo). 
The measured PE/MeV as a function of nominal energy for each event type for the two segments is also shown in Figure\,\ref{fig:spectra}.
The average light collection is 850$\mathrm{\pm20}$\,PE/MeV. 
Figure\,\ref{fig:nLi_fit} shows the (n,$^{6}$Li) peak fit to a Gaussian to extract the resolution. 
The (n,H) and $\mathrm{^{137}Cs}$ Compton edges were compared to simulation. 
Photon statistics are the dominant contribution to the width, followed by energy-independent geometric effects from the area mismatch between the square segment cross-section and round photocathode. 
Thus, the relative energy resolution of the 3 features are fit to the function $\sigma =  \sqrt{a^{2} + b^{2}/E}$ where $a$ represents the geometric term and $b$ the photon statistics term. 
The result is $\sigma$ = 4.0$\pm$0.2\% at 1\,MeV.

\subsection{Neutron characterization}

\begin{figure}[h!]
\centering
\includegraphics[width=0.49\textwidth]{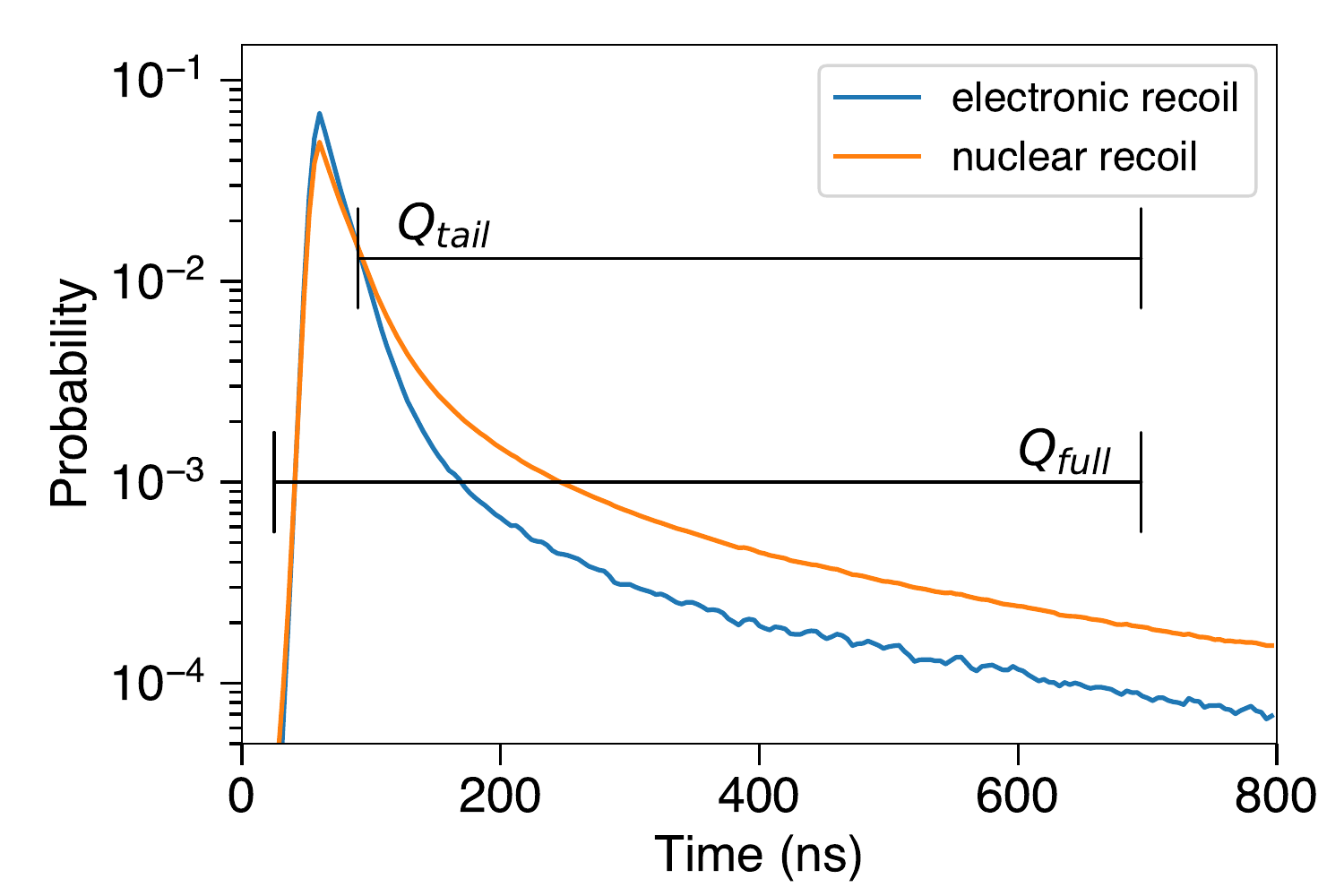} 
\includegraphics[width=0.49\textwidth]{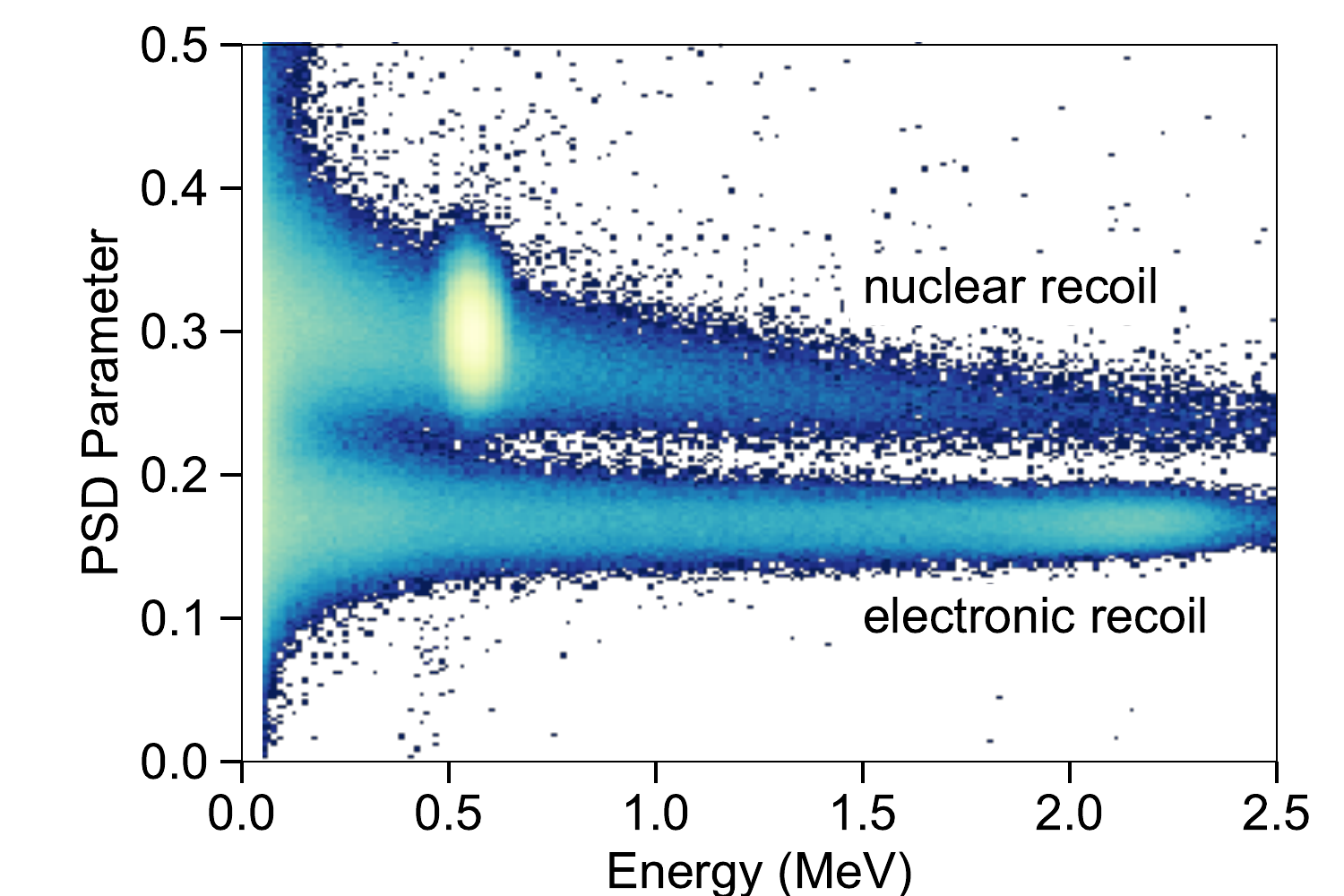} 
\caption{\textbf{(a)} Template waveforms from the PROSPECT-50 detector illustrating the difference between electronic and nuclear recoil energy depositions from a $^{252}$Cf source. The fraction of charge in the tail of the pulse with respect to the total integral charge is used as a discrimination metric. \textbf{(b)} PSD parameter distribution as a function of energy when exposed to a $^{252}$Cf neutron source. 
}
\label{fig:P50_2dPSD}
\end{figure}

A $\mathrm{^{252}Cf}$ neutron source positioned 1\,m from the center of PROSPECT-50 is used to study PSD performance and measure the neutron capture time in the $^{6}$LiLS. 
As specified in Equation\,\ref{eq:PSD}, $Q_{full}$ in PROSPECT-50 is defined by an integration window from 12\,ns before to 120\,ns after the half-height of the waveform's leading edge and $Q_{tail}$ as the charge integrated 40\,ns to 120\,ns after the leading edge half-height. 
Figure\,\ref{fig:P50_2dPSD} shows averaged waveforms for electronic recoil and nuclear recoil events and demonstrates the resulting photoelectron-weighted average PSD for the combined PMTs. 
Distinct electronic and nuclear recoil bands are observed between PSD parameters of [0.14, 0.20] and [0.24, 0.36], respectively, and a prominent peak at $\mathrm{\sim}$0.55\,MeV indicates neutron captures on $^{6}$Li.

\begin{figure}[h!]
\centering
\includegraphics[width=0.49\textwidth]{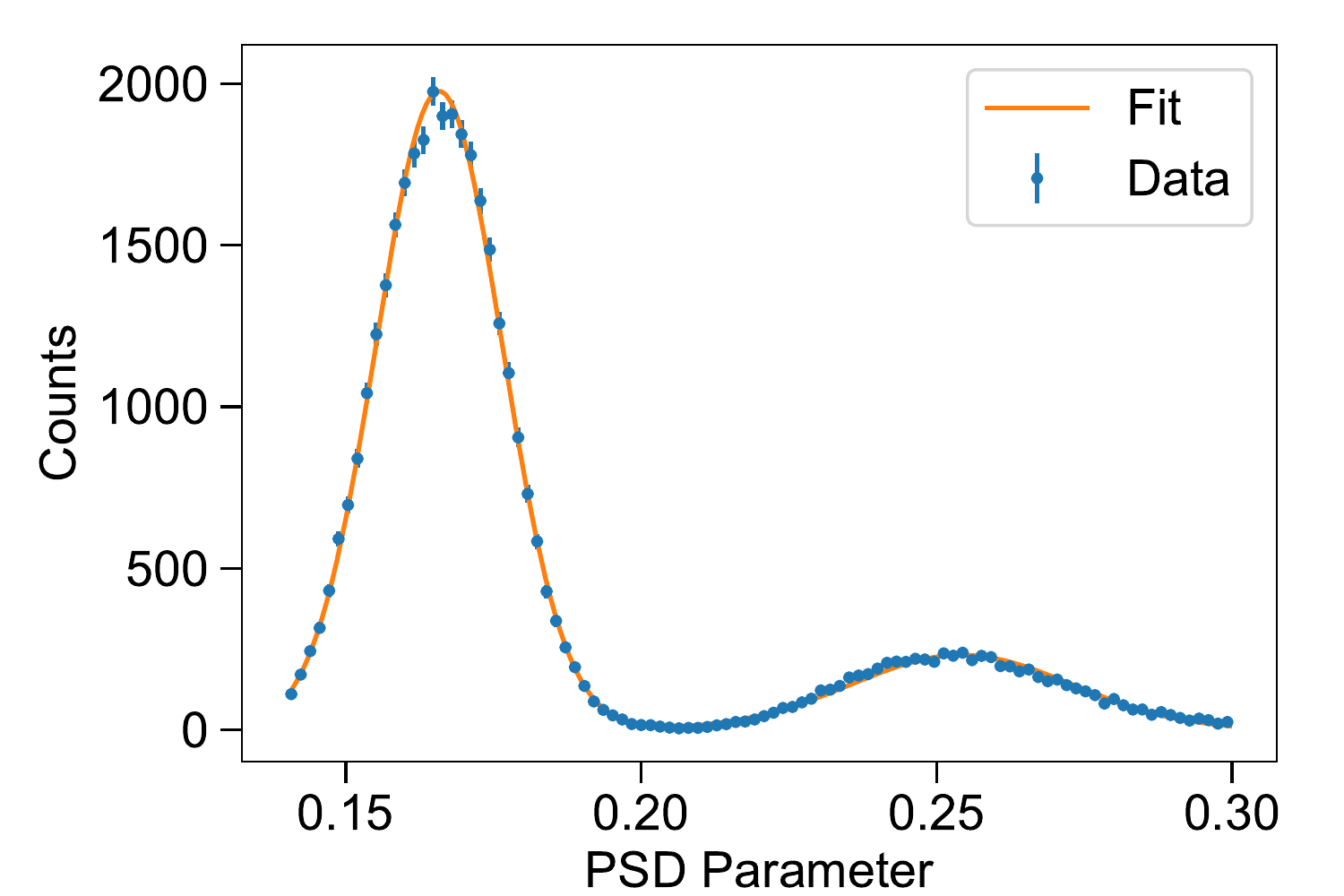} 
\includegraphics[width=0.49\textwidth]{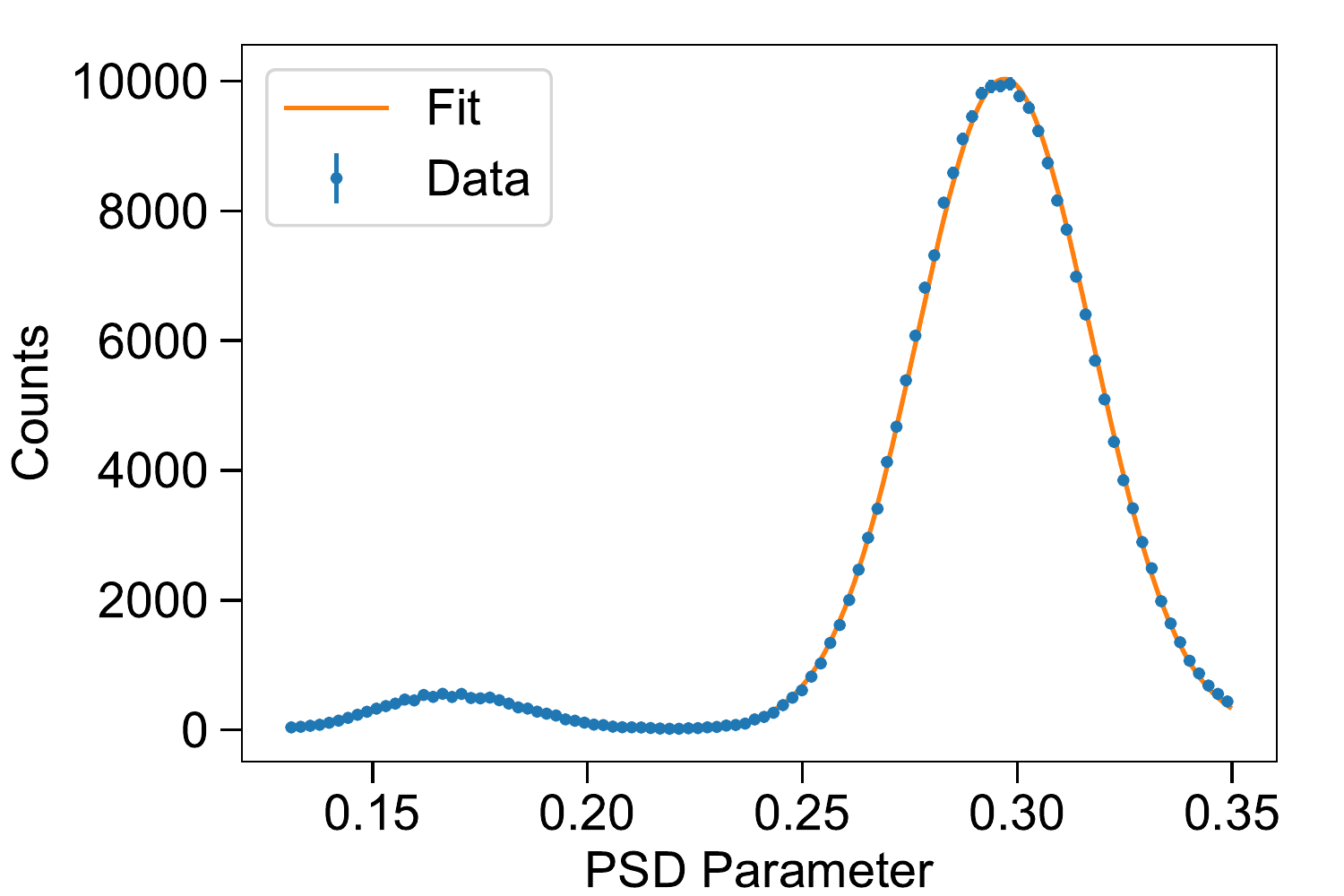} 
\caption{\textbf{(a)} Projections of PSD in IBD-like prompt space [$1 < E\mathrm{(MeV)} < 3$] and \textbf{(b)} in (n,$^{6}$Li) capture region [$0.4 < E\mathrm{(MeV)} < 0.7$]. 
}
\label{fig:P50_1dPSD}
\end{figure}

PSD projections, as illustrated in Figure\,\ref{fig:P50_1dPSD}, show the quality of separation between event types for ``prompt-like'' [$1 < E\mathrm{(MeV)} < 3$] as a substitute for IBD prompt positron events and the ``delay-like'' neutron capture [$0.4 < E\mathrm{(MeV)} < 0.7$] region of the spectrum. 
The separation figure-of-merit (FOM) is defined as

\begin{equation}
FOM = \frac{\vert\mu_{1}-\mu_{2}\vert}{FWHM_{1}+FWHM_{2}}
\end{equation}

\noindent where $\mu_{1}$, $\mu_{2}$ represent the means of the electronic and nuclear recoil distributions, respectively, and $FWHM_{1}$, $FWHM_{2}$ the full-width-half-maximum of the corresponding distribution. 
Here, both the prompt and delay PSDs exhibit FOM = 1.5, satisfying the PROSPECT requirement of FOM $>$1.

\begin{figure}[h!]
\begin{center}
\includegraphics[width=0.65\textwidth]{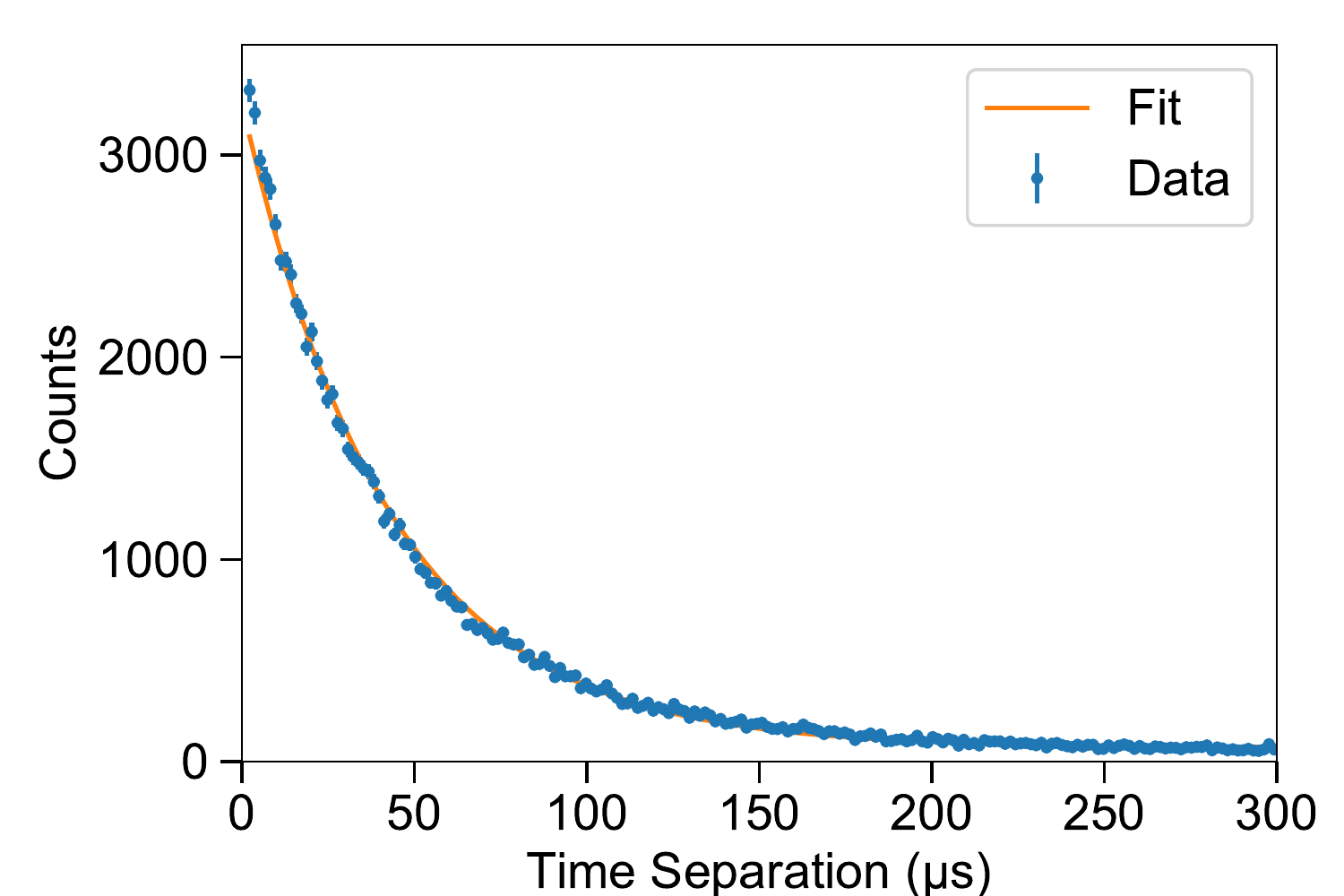}
\caption{Demonstration of the neutron capture time $\mathrm{\tau = 42.8\pm0.2\,\mu s}$  in the PROSPECT-50 detector.}
\label{fig:nLi_time}
\end{center}
\end{figure}

By identifying proton recoils that occur in a 300\,$\mathrm{\mu}$s window before a delayed neutron capture on $^{6}$Li, the characteristic neutron capture time ($\tau$) was measured for the PROSPECT-50 detector. 
The results are shown in Figure\,\ref{fig:nLi_time}. 
Fit with a simple exponential, $Ae^{(-t/\tau)}+C$, we obtain $\tau$ = 42.8$\pm$0.2\,$\mathrm{\mu s}$.
 
\section{Long-term stability}
\label{sec:stability}

The PROSPECT-50 detector was filled in July 2017 and has operated nearly continuously, providing a long term stability measure of a full production PROSPECT detector system.
Multiple calibrations have been performed and the PMT gains have been monitored using the OCS discussed in Section\,\ref{subsec:calibrationDesign}.
A mean single photoelectron value for each PMT is extracted from a fit of ambient background runs.
A sliding-window smoothing is applied to these data to reduce variation and produce a time-dependent gain correction for each PMT\footnote{The OCS ceased operation in early January, 2018, therefore PMT gain corrections for data after this time are not made.}. 

\begin{figure}[h] 
   \centering
   \includegraphics[width=.49\textwidth]{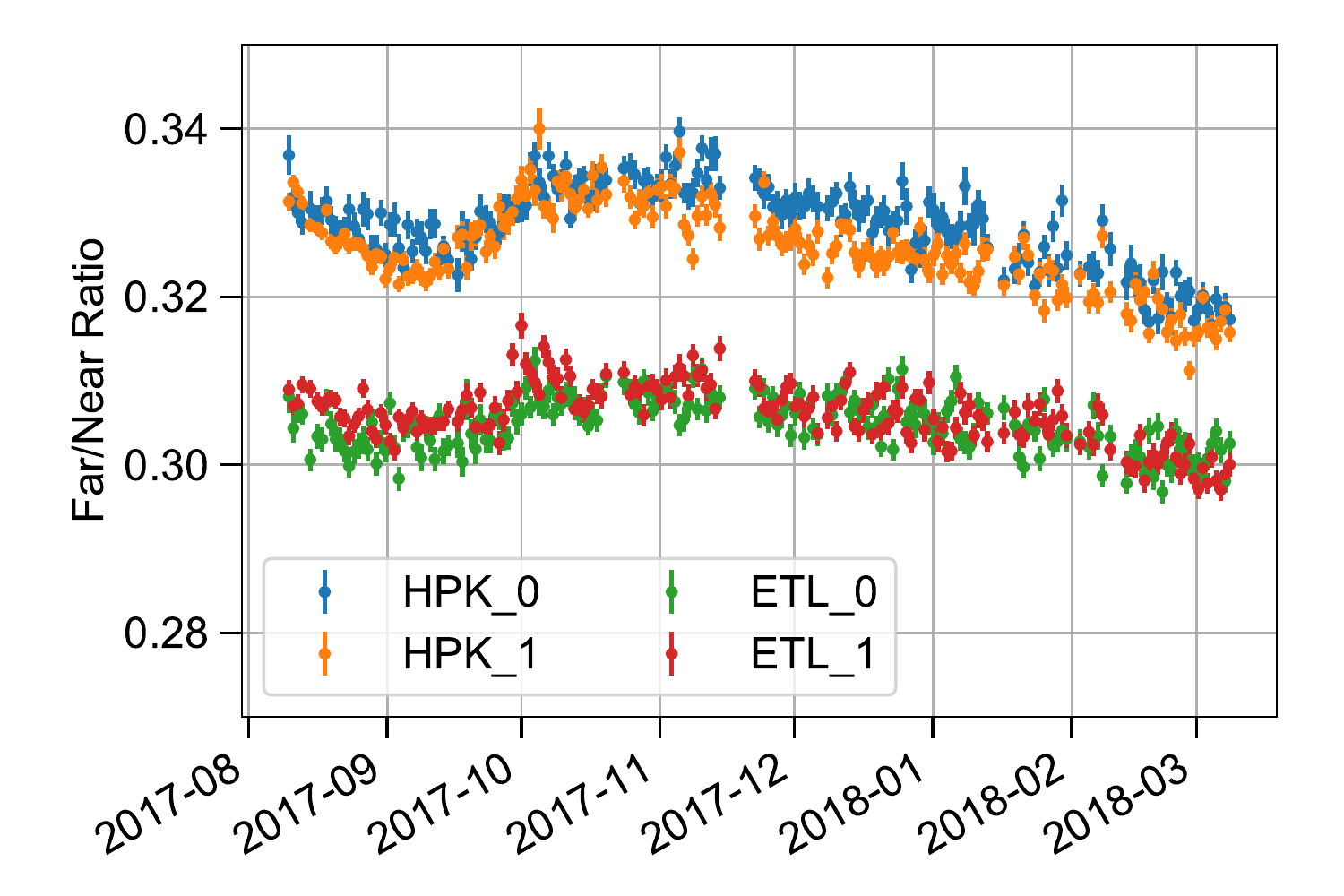}~
   \includegraphics[width=.49\textwidth]{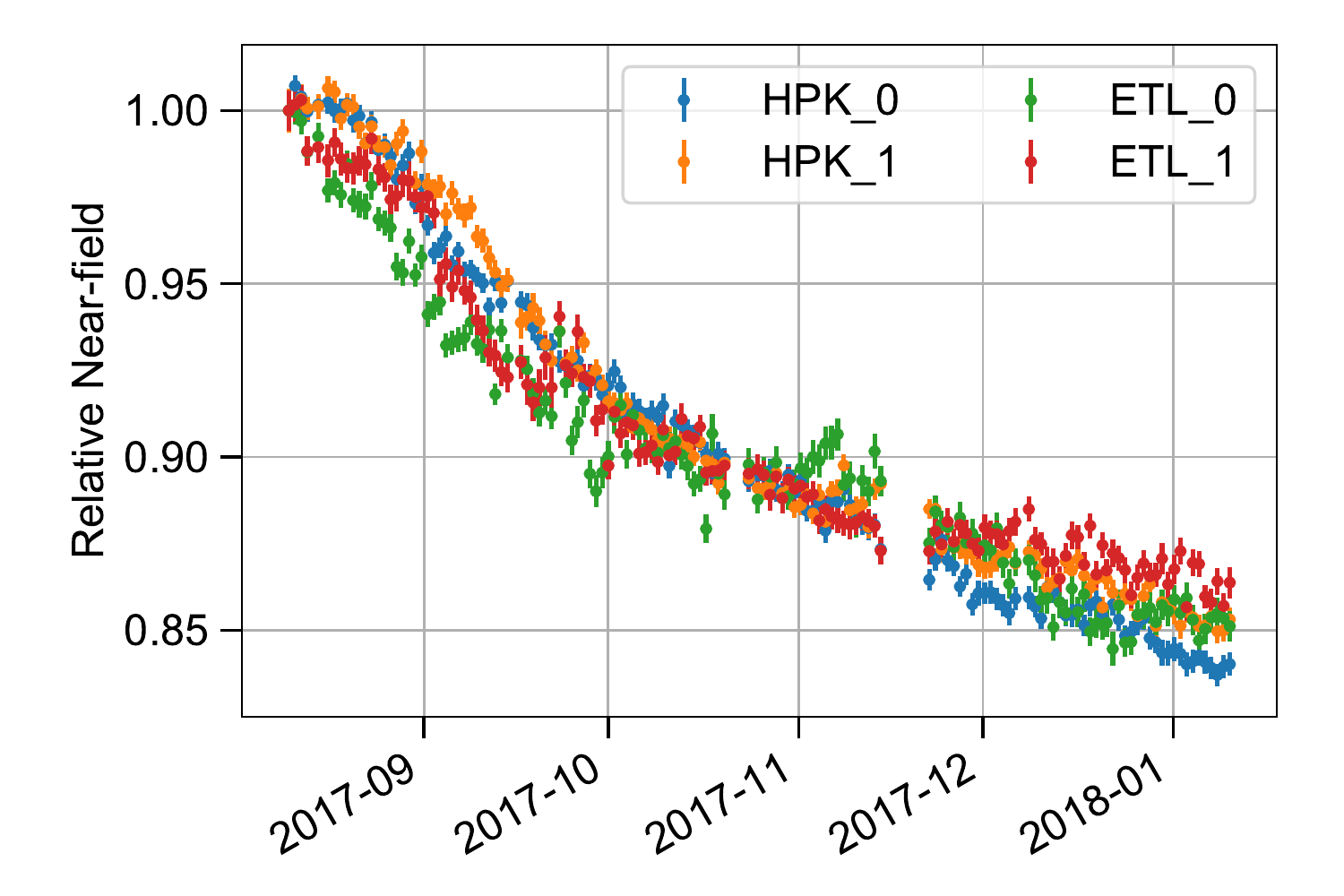} 
   \caption{\textbf{(a)} Ratio of fitted (n,$^{6}$Li) capture peaks far and near from each PMT. The two cells have different effective attenuation lengths based on the angular acceptance of the PMT reflector, but both show coherent changes of a few percent over seven months of operation.
   \textbf{(b)} Fitted (n,$^{6}$Li) capture peak locations for events near to each PMT, normalized to the first data-point and corrected for PMT gain-drift.}
   \label{fig:attenuation_length}
\end{figure}

The scintillator light yield and attenuation length are tracked separately over time using the point-like energy deposition from neutron captures on $^{6}$Li. 
Events are selected by their relative timing to have occurred either `near' or `far' from one of the PMTs.
A 2\,ns wide window selects events within $\sim$20\,cm of the segment end.
The light yield of the scintillator is strongly correlated with the near-field collection, while the ratio of the far-field to near-field response to neutron captures provides a measure of the effective attenuation.
Figure\,\ref{fig:attenuation_length} shows the measured far-to-near ratio for all four PMTs.
This ratio has varied by approximately 3\% for each cell over seven months of observation. 

\noindent Figure\,\ref{fig:attenuation_length} shows the relative fitted peak location of (n,$^{6}$Li) for events located near each PMT.
A 15\% decrease in the light collection is observed, consistent across all PMTs\footnote{Small gaps in the data correspond to periods of special calibrations or temporary pauses in DAQ operation.}.
A corresponding decrease in the fraction of late-light ($Q_{tail}$) of signals has also been observed.
These two effects are strong indicators of oxygen quenching\,\cite{OKeeffe:2011dex,Li:2011a} and no external environmental variables were found to be correlated with the data. 
In September 2017 the detector enclosure was opened and a sample of $^{6}$LiLS was removed. 
Measurements of this sample's light yield were made before and after bubbling with nitrogen.
A $\sim$10\% increase in light output was observed along with an increase in the $Q_{tail}$ fraction.
This implicates oxygen quenching as a likely explanation for the reduced light yield in PROSPECT-50.

Due to the design of this prototype, PROSPECT-50 was filled with scintillator without the secondary containment lid sealed.
Though an increased flow of cover gas was employed, it is believed that insufficient purging may have contributed to oxygen quenching of the scintillator. 
This, combined with possible outgassing of the acrylic support structure, has produced a non-optimal environment for the $^{6}$LiLS.
Despite this reduction of light collection, PROSPECT-50 maintains a PSD figure of merit greater than 1.0 and an energy resolution below 4.5\% at 1\,MeV.


\section{Conclusions}
\label{sec:conclusions}

Using a full-scale prototype for PROSPECT we have demonstrated the performance of a segmented $^{6}$Li-loaded liquid scintillator detector. 
We measure an excellent light collection of 850$\pm$20~PE/MeV and energy resolution of $\sigma$ = 4.0$\pm$0.2\% at 1\,MeV with two 117.6$\times$14.5$\times$14.5~cm$^{3}$ detector segments. 
The ability to efficiently separate high $dE/dx$ (electronic recoil) and low $dE/dx$ (nuclear recoil) depositions and distinguish neutron captures on $^{6}$Li is demonstrated. 
The effective scintillation attenuation length, position reconstruction, and neutron capture time are also measured in this geometry. 
The long-term stability of the scintillator has been monitored and the observations are consistent with oxygen quenching. 
The presented results meet the specifications for identifying neutrino events and rejecting backgrounds for the PROSPECT experiment and may have significant value in other applications, such as efficient neutron detection. 

\section{Acknowledgements}
This material is based upon work supported by the following sources: US Department of Energy (DOE) Office of Science, Office of High Energy Physics under Award No. DE-SC0016357 and DE-SC0017660 to Yale University, under Award No. DE-SC0017815 to Drexel University, under Award No. DE-SC0008347 to Illinois Institute of Technology, under Award No. DE-SC0016060 to Temple University, under Contract No. DE-SC0012704 to Brookhaven National Laboratory, and under Work Proposal Number  SCW1504 to Lawrence Livermore National Laboratory. This work was performed under the auspices of the U.S. Department of Energy by Lawrence Livermore National Laboratory under Contract DE-AC52-07NA27344 and by Oak Ridge National Laboratory under Contract DE-AC05-00OR22725. This work was also supported by the Natural Sciences and Engineering Research Council of Canada (NSERC) Discovery program under grant \#RGPIN-418579 and Province of Ontario.

Additional funding was provided by the Heising-Simons Foundation under Award No. \#2016-117 to Yale University. J.G. is supported through the NSF Graduate Research Fellowship Program and A.C. performed work under appointment to the Nuclear Nonproliferation International Safeguards Fellowship Program sponsored by the National Nuclear Security Administration's Office of International Nuclear Safeguards (NA-241).

We further acknowledge support from Yale University, the Illinois Institute of Technology, Temple University, Brookhaven National Laboratory, the Lawrence Livermore National Laboratory LDRD program, the National Institute of Standards and Technology, and Oak Ridge National Laboratory. We gratefully acknowledge the support and hospitality of the High Flux Isotope Reactor and Oak Ridge National Laboratory, managed by UT-Battelle for the U.S. Department of Energy.

\newpage
\bibliographystyle{./style/h-physrev}

\bibliography{p50}

\begin{thebibliography}{10}
\bibitem{disclaimer}
Certain trade names and company products are mentioned in the text or
  identified in illustrations in order to adequately specify the experimental
  procedure and equipment used. In no case does such identification imply
  recommendation or endorsement by the National Institute of Standards and
  Technology, nor does it imply that the products are necessarily the best
  available for the purpose.
\end{thebibliography}

\end{document}